\documentclass[fleqn,usenatbib]{mnras}

\usepackage{newtxtext,newtxmath}

\usepackage[T1]{fontenc}

\DeclareRobustCommand{\VAN}[3]{#2}
\let\VANthebibliography\thebibliography
\def\thebibliography{\DeclareRobustCommand{\VAN}[3]{##3}\VANthebibliography}

\usepackage[dvipdfmx]{graphicx}	
\usepackage{amsmath}	
\usepackage{breqn}
\usepackage{multirow}

\newcommand{\Mh}{M_{\rm h}}
\newcommand\Msun{M_{\odot}}
\newcommand\Mmin{M_{\min}}
\newcommand\Mstar{M_{\star}}
\newcommand\Meff{M_{\rm eff}}

\newcommand\fsat{f_{\rm sat}}
\newcommand\ffake{f_{\rm fake}}
\newcommand\bg{b_{\rm g}}
\newcommand\bh{b_{\rm h}}
\newcommand\ngal{\bar{n}_{\rm ELG}}
\newcommand\sigmalogM{\sigma_{\log{L}}}

\newcommand{\Nc}{N_{\rm c}}
\newcommand{\Ns}{N_{\rm s}}
\newcommand{\Ntot}{N_{\rm tot}}

\newcommand{\NtotELG}{N_{\rm tot, ELG}}
\newcommand{\oii}{[\ion{O}{II}]}

\newcommand{\Lth}{L_{\rm th}}
\newcommand{\Mpivot}{M_{\rm pivot}}
\newcommand{\Mtrans}{M_{\rm trans}}
\newcommand{\Lzero}{L_{\rm 0}}
\newcommand{\fmax}{f_{\rm ELG}^{\rm max}}
\newcommand{\fmin}{f_{\rm ELG}^{\rm min}}

\title[New HOD modelling with LF constraint for ELGs]{
A new constraint on galaxy--halo connections of [\ion{O}{II}] emitters via HOD modelling with angular clustering and luminosity functions from the Subaru HSC survey}

\author[Shogo Ishikawa et al.]{
Shogo Ishikawa$^{1,2,3,4}$\footnotemark{\thanks{shogo.ishikawa.astro@gmail.com, shogo.ishikawa@yukawa.kyoto-u.ac.jp}},
Teppei Okumura$^{5,6}$,
Masao Hayashi$^{4}$,
and 
Tsutomu T. Takeuchi$^{7,8}$
\\
$^{1}$Department of Liberal Arts and Basic Sciences, College of Industrial Technology, Nihon University, Narashino, Chiba 275-8576, Japan\\
$^{2}$Center for Gravitational Physics and Quantum Information, Yukawa Institute for Theoretical Physics, Kyoto University, Sakyo-ku, Kyoto 606-8502, Japan\\
$^{3}$Center for Computational Astrophysics, National Astronomical Observatory of Japan, Mitaka, Tokyo 181-8588, Japan\\
$^{4}$National Astronomical Observatory of Japan, Mitaka, Tokyo 181-8588, Japan\\
$^{5}$Institute of Astronomy and Astrophysics, Academia Sinica, No.~1, Section~4, Roosevelt Road, Taipei 10617, Taiwan\\
$^{6}$Kavli Institute for the Physics and Mathematics of the Universe (WPI), UTIAS, The University of Tokyo, Kashiwa, Chiba 277-8583, Japan\\
$^{7}$Division of Particle and Astrophysical Science, Nagoya University, Furo-cho, Chikusa-ku, Nagoya 464-8602, Japan\\
$^{8}$The Research Center for Statistical Machine Learning, The Institute of Statistical Mathematics, 10-3 Midori-cho, Tachikawa, Tokyo 190-8562, Japan
}

\date{Accepted XXX. Received YYY; in original form ZZZ}

\pubyear{2024}

\begin{document}
\label{firstpage}
\pagerange{\pageref{firstpage}--\pageref{lastpage}}
\maketitle

\begin{abstract}
Establishing a robust connection model between emission-line galaxies (ELGs) and their host dark haloes is of paramount importance in anticipation of upcoming redshift surveys. 
We propose a novel halo occupation distribution (HOD) framework that incorporates galaxy luminosity, a key observable reflecting ELG star-formation activity, into the galaxy occupation model. 
This innovation enables prediction of galaxy luminosity functions (LFs) and facilitates joint analyses using both angular correlation functions (ACFs) and LFs. 
Using physical information from luminosity, our model provides more robust constraints on the ELG--halo connection compared to methods relying solely on ACF and number density constraints. 
Our model was applied to $\oii$-emitting galaxies observed at two redshift slices at $z=1.193$ and $1.471$ from the Subaru Hyper Suprime-Cam PDR2. 
Our model effectively reproduces observed ACFs and LFs observed in both redshift slices. 
Compared to the established \citeauthor{geach12} HOD model, our approach offers a more nuanced depiction of ELG occupation across halo mass ranges, suggesting a more realistic representation of ELG environments. 
Our findings suggest that ELGs at $z\sim1.4$ may evolve into Milky-Way-like galaxies, as their inferred halo masses evolve accordingly based on the extended Press--Schechter formalism, highlighting their role as potential building blocks in galaxy formation scenarios. 
By incorporating the LF as a constraint linking galaxy luminosity to halo properties, our HOD model provides a more precise understanding of ELG-host halo relationships. 
Furthermore, this approach facilitates the generation of high-quality ELG mock catalogues of for future surveys. 
As the LF is a fundamental observable, our framework is potentially applicable to diverse galaxy populations, offering a versatile tool for analysing data from next-generation galaxy surveys. 
\end{abstract}




\section{Introduction} \label{sec:intro}
The mass of dark matter haloes is one of the most fundamental quantities for characterising galaxy populations in the $\Lambda$-dominated cold dark matter ($\Lambda$CDM) paradigm \citep[e.g.,][]{white78,white91}. 
The halo occupation distribution (HOD) model provides a straightforward connection between observed galaxies and invisible background matter fluctuation \citep[e.g.,][]{berlind02,berlind03,vdbosch07}. 
Luminous red galaxies (LRGs) are thought to be old and massive quiescent galaxies, and many galaxy surveys have extensively targeted LRGs as a tracer of the large-scale structure, such as the Sloan Digital Sky Survey \citep[SDSS;][]{eisenstein01}, the 2SLAQ \citep{cannon06}, and the Dark Energy Spectroscopic Instrument \citep[DESI;][]{zhou23}. 
The standard HOD model proposed by \citet{zheng05}, characterised by stochastic central occupation with a Poisson satellite occupation model, has accurately reproduced the observed projected correlation functions of LRGs \citep[e.g.,][]{zhou21,ishikawa21,ishikawa24}, and has also been widely applied to interpret LRG clustering measurements \citep[e.g.,][]{matsuoka11}. 
The findings were further supported by cosmological hydrodynamical simulations that confirmed the expected number of LRG-like red and luminous galaxies as a function of dark halo mass \citep{bose19,beltz20}. 

In addition to their success in characterising LRGs, emission-line galaxies (ELGs) have become key targets in recent extensive galaxy surveys. 
Ionised gas is produced in the vicinity of young, massive stars. 
Then, through the processes of electron--ion recombination and energy level transitions from exited states through electric multipole radiation, strong forbidden and permitted lines are produced from the gaseous nebulae \citep[e.g.,][]{anders03,byler17}. 
Young galaxies with high star-formation activities tend to exhibit such emission lines in their spectral energy distributions, making them identifiable as ELGs through narrow-band photometry that captures strong emission lines compared to the spectral continuum. 

Since ELGs with identical emission lines observed through a single narrow-band filter are almost free from photometric redshift uncertainties, extensive narrow-band photometric surveys have played a crucial role in identifying and cataloguing these galaxies. 
Surveys such as COSMOS \citep{laigle16,saito20}, HiZELS \citep{geach08}, and the Subaru Hyper Suprime-Cam Subaru Strategic Programme \citep[HSC SSP;][]{aihara18} have successfully constructed tomographic ELG catalogues spanning broad redshift ranges. 
Narrow-band surveys efficiently identify galaxies with strong emission lines, particularly useful for constructing tomographic ELG catalogues. 
These catalogues enable statistical studies of galaxy clustering and evolution by providing large samples of ELGs with minimal redshift uncertainties. 

However, it remains unclear if ELGs follow the same halo occupation model as LRGs. 
Unlike LRGs, ELGs reside in diverse environments and exhibit ongoing star formation, necessitating a unique HOD framework that accounts for their distinct halo occupation patterns. 
It is necessary to develop alternative HOD models tailored to current and upcoming ELG surveys, such as the Subaru Prime Focus Spectrograph Survey \citep[PFS;][]{takada14} and the DESI \citep{desi16}. 

\citet{geach12} proposed an HOD model specifically for ELGs, using semi-analytic galaxy formation models \citep[GALFORM;][]{cole00}, to explain the clustering and angular distribution of H$\alpha$-emitting galaxies at $z=2.23$. 
Their model incorporates two components of central ELGs: one Gaussian-distributed in less massive haloes and another following an error function in more massive haloes. 
This flexible model, with nine free parameters, has successfully explained the clustering and distribution of not only H$\alpha$ emitters but also other ELGs \citep[e.g.,][]{cochrane18,okumura21,gao22}. 
While this model has proven effective in reproducing observed clustering, the direct link between ELG properties and halo characteristics remains incomplete. 

In recent years, significant progress has been made in understanding the connection between ELGs and their host dark haloes using cosmological hydrodynamical simulations, in preparation for forthcoming extensive galaxy redshift surveys. 
For instance, \citet{osato23} reported further support of the \citet{geach12} model for ELG-like galaxies rather than the \citet{zheng05} model by directly counting the expectation number as a function of halo masses selected in the IllustrisTNG simulation \citep{springel18}. However, there are some discrepancies in reproducing the ACFs from the directly counted HOD of ELGs, which indicates that improvements are still needed. 
Furthermore, \citet{hadzhiyska21} also constructed mock ELG catalogues using the subhalo catalogue of the IllustrisTNG by applying the stellar population synthesis model different from that used by \citet{osato23}. They then found that the directly counted ELG occupations are still different from those prodicted by \citet{geach12}. 
Therefore, given the imminent arrival of large ELG datasets from next-generation galaxy surveys, it is crucial to revisit the HOD framework for ELGs. 
Constructing a robust HOD model not only aids in interpreting observed ELG clustering but also in generating accurate mock catalogues for future studies. 

In this paper, we propose a novel HOD framework that directly incorporates ELG luminosities, a key observable that directly reflects the star-formation activity of ELGs, into the halo occupation statistics. 
By linking ELG line luminosities to dark halo masses, our HOD model provides stronger constraints on the ELG--dark halo connection. 
The central ELG occupation in our model primarily consists of an error function-based occupation with a decaying component at the massive end. 
This simplified model offers a more straightforward interpretation of observed ELG characteristics compared to HOD analyses that rely on more complex occupation functions. 
Compared to the widely adopted model of \citet{geach12} for interpreting the ELG clustering, our framework integrates luminosity functions directly into the HOD modelling, successfully establishing a direct connection between observable baryonic properties and dark matter density fluctuation.  
Consequently, our model provides improved predictions for the spatial distribution of ELGs, making it particularly well-suited for application to upcoming extensive galaxy surveys. 

During the early stages of establishing the HOD model, a method called the conditional luminosity function \citep[CLF;][]{yang03,vdbosch03a,vdbosch03b} model was already proposed to link galaxy occupation with galaxy luminosity functions. 
The CLF model has the advantage of constraining the differential number density of galaxies within the luminosity range of $\pm dL$ with a halo mass of $\Mh$, represented as $\langle N \rangle (\Mh) = \int dL \Phi(L|\Mh)L$, where $\langle N \rangle$ is the expected number of galaxies, and $\Phi$ represents the luminosity function of galaxies. 
However, the CLF assumes a functional form for the luminosity function, in many cases adopting the Schechter function \citep{schechter76}. 
Recent observational results have indicated the possibility of an excess in the bright-end of luminosity functions in high-redshift Universe, for instance, the various types of ELGs \citep{hayashi20} and Lyman break galaxies \citep{harikane22}. 
In our approach, we do not assume any functional forms for galaxy luminosity functions, enabling to flexibly constrain the ELG--halo connection. 

This paper is organised as follows: In Section~\ref{sec:observation}, we present the observational parameters of the HSC SSP Deep/UltraDeep layers, detail our ELG catalogues obtained from these layers, and calculate the angular correlation functions and luminosity functions of our ELG samples, including estimates of covariance matrices. 
In Section~\ref{sec:hod_model}, we introduce our HOD model, which incorporates line luminosity in the central galaxy occupation model. 
Our proposed HOD model is then used to perform HOD-model analyses on our $\oii$-emitting galaxy samples, and the results are compared with those derived using the HOD model by \citet{geach12}, which is widely used to interpret the galaxy--halo connection for ELGs, in Section~\ref{sec:hod}. 
Section~\ref{sec:discussion} discusses the evolutionary link between our $\oii$ samples and low-$z$ galaxy populations, as well as the $\oii$ emitters--host halo connection based on the results obtained in the previous section. 
We conclude and summarise this study in Section~\ref{sec:conclusion}. 

Throughout this paper, we employ the flat $\Lambda$CDM cosmological parameters constrained by observations of the cosmic microwave background radiation from the Planck satellite \citep{planck18}. 
The Planck 2018 cosmological parameters are as follows: the cosmic density parameters are $\Omega_{\rm m} = 0.3153$, $\Omega_{\rm \Lambda} = 0.6847$, and $\Omega_{\rm b} = 0.0493$; the dimensionless Hubble parameter is $h = 0.6736$; the matter fluctuation normalised over $8 h^{-1}$ Mpc is $\sigma_{8} = 0.8111$; and the primordial spectral index is $n_{\rm s} = 0.9649$. 
Stellar and halo masses are denoted by $\Mstar$ and $\Mh$, and are scaled in units of $\Msun$ and $h^{-1}\Msun$, respectively. 
All magnitudes are given in AB magnitude. 
The logarithm denoted as $\ln$ refers to the natural logarithm, while all other logarithms in this paper are base $10$. 

\section{Observations} \label{sec:observation}
\subsection{Photometric data} \label{subsec:data}
The ELGs used in this paper are selected from photometric data obtained by the HSC SSP \citep{aihara18}. 
The HSC SSP is an extensive multi-wavelength photometric survey that fully utilises the unique capability of the wide-field imaging of the Hyper Suprime-Cam \citep[HSC;][]{miyazaki18,komiyama18,furusawa18}, which has a $1.5$ degree diameter field-of-view and is installed at the prime focus of the Subaru Telescope. 
Our ELG samples are obtained from the HSC SSP public data release 2 \citep[PDR2;][]{aihara19}. 

The HSC SSP consists of three distinct layers: Wide, Deep, and UltraDeep (UD), all observed using broad-band (BB) filters of varying depths. 
The BB filters of the HSC SSP include optical $g$-, $r$-, $i$-, $z$-, and $y$-band filters \citep{kawanomoto18}. 
In addition to the BB observations, the Deep and UD layers are also observed with four narrow-band (NB) filters: NB387, NB816, NB921, and NB1010. 
In this study, we only use two NB datasets, NB816 and NB921, which cover almost the entire regions of the Deep and UD layers. 
The central wavelength and width of the NB transmission curve for NB816 (NB921) are $8,177~\text{\AA}$ ($9,214~\text{\AA}$) and $113~\text{\AA}$ ($135~\text{\AA}$), respectively. 
For more detailed information about the photometric data of the HSC SSP PDR2, refer to \citet{aihara19}. 

\subsection{Emission-line galaxy catalogue} \label{subsec:catalogue}
\begin{table*}
\caption{Details of physical properties of the HSC SSP $\oii$-emitting galaxy samples}
\label{tab:elg}
\begin{tabular}{ccccccccc} \hline \hline
NB filter & Effective redshift & Redshift range & Area [deg$^{2}$] & $N_{\rm g}^{a}$ & $n_{\rm g}$ $[({\rm cMpc}/h)^{-3}]^{b}$ & $m_{\rm th}$ [mag]$^{c}$ & $f_{\rm th}$ [erg/s/cm$^{2}$]$^{d}$ & $L_{\rm th}$ [erg/s]$^{e}$ \\ \hline
NB816 & $1.193$ & $1.178 - 1.208$ & $16.284$ & $8,267$ & $5.50 \times 10^{-3}$ & $23.75$ & $3.0 \times 10^{-17}$ & $2.6 \times 10^{41}$ \\
NB921 & $1.471$ & $1.453 - 1.489$ & $16.864$ & $7,780$ & $3.68 \times 10^{-3}$ & $23.75$ & $3.0 \times 10^{-17}$ & $4.3 \times 10^{41}$ \\ \hline
\multicolumn{5}{l}{\footnotesize$^a$ the number of ELG samples. }\\
\multicolumn{5}{l}{\footnotesize$^b$ the number density of ELG samples. }\\
\multicolumn{5}{l}{\footnotesize$^c$ threshold magnitude in cmodel manigutde. }\\
\multicolumn{5}{l}{\footnotesize$^d$ threshold $\oii$ line flux. }\\
\multicolumn{5}{l}{\footnotesize$^e$ threshold $\oii$ line luminosity. }\\
\end{tabular}
\end{table*}

\citet{hayashi20} constructed ELG catalogues at $z<1.6$ using photometric data from the HSC SSP observation. 
In this paper, we focus on $\oii$-emitting galaxies at two redshift slices selected by \citet{hayashi20}, enabling us to demonstrate our HOD model and reveal the relationship between ELGs and host haloes. 
One ELG sample is at redshift $z=1.193$, observed by their emission lines with NB816, and the other is at $z=1.471$, observed with NB921. 
Both populations exhibit small redshift ranges, typically $\Delta z \sim 0.03$.

To ensure the homogeneity of our ELG samples across the survey fields, we set magnitude and line luminosity thresholds for each ELG sample. 
These thresholds are similar to those applied to the $\oii$-emitting galaxies of \citet{okumura21}, who conducted clustering and HOD analyses using the same $\oii$-emitter samples as those used in this paper. 
We set the cmodel magnitude\footnote{The cmodel magnitudes are defined as a linear combination of a de Vaucouleurs and an exponential model of source objects, and are suitable for extended sources \citep{stoughton02,abazajian04}.} threshold as $m_{\rm X} \leq 23.50$, where X is either NB816 or NB921, and the corresponding line flux threshold as $f_{\rm X} \geq 3.0 \times 10^{-17}$ [erg/s/cm$^{2}$] to maintain a galaxy number count completeness of more than 70$\%$ at the faint end. 
These thresholds are fully consistent with \citet{okumura21}. 
Assuming the Planck 2018 cosmologies and effective redshifts, the corresponding line luminosity thresholds of our $\oii$ emitter samples observed through NB816 and NB921 filters are $2.6 \times 10^{41}$ [erg/s] and $4.3 \times 10^{41}$ [erg/s], respectively. 

We summarise the properties of our ELG samples in Table~\ref{tab:elg}. 
Note that the effective areas of each NB observation are calculated after masking bright stars and regions at the survey edges. 
For more information about the photometric data of NB observation by the HSC SSP and the ELG catalogue construction, refer to \citet{hayashi20}. 

Although our ELG samples were carefully selected to maintain high completeness ($>70~\%$), some observational limitations and selection biases remain. 
These include contamination from misidentified emission lines originating from galaxies at unintended redshifts, which we explicitly accounted for by introducing a contamination fraction function in our HOD model (Section~\ref{subsubsec:fake_elg_fraction}). 
Additionally, uncertainties in intrinsic luminosity estimates due to dust-extinction corrections, calibrated using Balmer decrements derived from SDSS galaxies, may affect our luminosity-based constraints (Section~\ref{subsec:lf}). 
Finally, despite our relatively large survey area ($\sim 16~{\rm deg}^{2}$), cosmic variance could slightly influence clustering measurements; we mitigated this effect by employing jackknife resampling (Section~\ref{subsection:acf_and_cov}, \ref{subsec:lf}). 
We emphasise, however, that these observational uncertainties and biases have been carefully incorporated into our analysis, and thus do not significantly impact our main results. 

\subsection{Angular correlation functions} \label{subsection:acf_and_cov}
The angular two-point correlation functions (hereafter ACFs) represent the excess probability of finding galaxy pairs compared to a random distribution as a function of the separation scale of galaxy pairs \citep{totsuji69,peebles80}. 
To calculate the ACFs of ELGs from the HSC SSP observation, we employ the well-known estimator proposed by \citet{landy93} as follows:
\begin{equation}
\omega(\theta) = \frac{{\rm DD}(\theta) - 2{\rm DR}(\theta) + {\rm RR}(\theta)}{{\rm RR}(\theta)},
\label{eq:ls93}
\end{equation}
where $\omega(\theta)$ is the ACF at the separation angle $\theta$, and ${\rm DD}(\theta)$, ${\rm DR}(\theta)$, and ${\rm RR}(\theta)$ denote the pair counts of galaxy--galaxy, galaxy--random, and random--random points with separation angle $\theta$, respectively. 
We adopt the angular separation scale as $-3.4 \le \log_{10}(\theta) \le 0.0$ with a bin size of $\log_{10}(\delta \theta) = 0.1$ in degrees, and the pair counts are conducted within a separation angle scale of $\theta \pm \delta \theta$. 
The random points are distributed over the same survey footprints of the HSC SSP after masking, and the number of random points is approximately $100$ times that of the observed $\oii$-emitter samples to reduce Poisson noise. 

Covariance matrices represent the correlations of signals across the bins of a dataset, and reliable covariance matrices are essential for accurate parameter estimation \citep[e.g.,][]{norberg09,ishikawa24}. 
We evaluate the covariance matrices of the observed ACFs of $\oii$-emitting galaxies using the delete-one jackknife (hereafter, JK) resampling method, which divides the entire region into subregions and calculates the ACFs by deleting one subregion, repeating this procedure for all subregions \citep{shao89,norberg09}. 
We employ the \textsc{kmeans_radec} package\footnote{https://github.com/esheldon/kmeans\_radec}, which is based on the k-means tree algorithm, to divide the survey footprint into subregions with similar areas. 
The HSC SSP Deep layers are divided into $82$ $(85)$ subregions for $\oii$-emitters observed through the NB816 (NB921) filter. 
The $(i, j)$ element of the covariance matrix measured in the $k$-th subfield, $C_{k, ij}$, is computed as follows:
\begin{equation}
C_{k, ij} = \frac{N-1}{N} \sum_{\ell=1}^{N} \left( \omega_{\ell} \left(\theta_{i} \right) - \bar{\omega} \left(\theta_{i} \right) \right) \times \left( \omega_{\ell} \left(\theta_{j} \right) - \bar{\omega} \left(\theta_{j} \right) \right), 
\label{eq:jk}
\end{equation}
where $N$ denotes the number of subregions, and $\omega_{\ell}$ and $\bar{\omega}$ represent the $\ell$-th JK realisation and mean ACFs over entire JK realisations, respectively. 

Note that the HSC SSP Deep layers consist of four distinct fields, and the ACFs and their covariance matrices for the entire regions are evaluated by weighting these quantities from each field. 
Following the procedure outlined in \citet{okumura21}, the weighted average ACFs of the $i$-th angular bin and the covariance matrices of the $(i, j)$ element, i.e., $\omega \left( \theta_{i} \right)$ and $C_{ij}$, can be calculated using those measured in the $k$-th subfield, denoted as $\omega_{k} \left( \theta_{i} \right)$ and $C_{k, ij}$, respectively:
\begin{equation}
\omega \left( \theta_{i} \right) = \frac{\sum_{k=1}^{4} W_{k}^{2} \left( \theta_{i} \right)  \omega_{k} \left( \theta_{i} \right)}{\sum_{k=1}^{4} W_{k}^{2} \left( \theta_{i} \right)}, 
\label{eq:weighted_acf}
\end{equation}
and 
\begin{equation}
C_{ij} = \frac{\sum_{k=1}^{4} W_{k}^{2} \left( \theta_{i} \right) W_{k}^{2} \left( \theta_{j} \right) C_{k, ij}}{\sum_{k=1}^{4} W_{k}^{2} \left( \theta_{i} \right) W_{k}^{2} \left( \theta_{j} \right)}. 
\label{eq:weighted_cov}
\end{equation}
Here, $W_{k} \left( \theta_{i} \right)$ represents the inverse variance of the $i$-th angular bin, corresponding to the square root of the $i$-th diagonal element of the covariance matrix, measured in the $k$-th subfield, i.e., 
\begin{equation}
W_{k} \left( \theta_{i} \right) = \frac{1}{\sqrt{C_{k, ii}}}.
\end{equation}

\subsection{Luminosity functions} \label{subsec:lf}
The luminosity functions of galaxies (hereafter LFs) describe the differential number density of galaxies measured in comoving volumes as a function of galaxy luminosities \citep[e.g.,][]{hubble31,abell77,binggeli88,favole24}. 
\citet{schechter76} proposed an analytical formula to express the observed LFs of local galaxies and cluster member galaxies with luminosities between $L$ and $L + dL$, $\Phi(L) dL$, as follows:
\begin{equation}
\Phi(L) dL = \left( \frac{\phi^{\star}}{L^{\star}} \right)  \left( \frac{L}{L^{\star}} \right)^{\alpha} \exp{\left( -\frac{L}{L^{\star}} \right)} dL,
\label{eq:lf}
\end{equation}
where $\phi^{\star}$ and $L^{\star}$ represent the characteristic number density and luminosity of galaxies, and $\alpha$ is the power-law slope of the LFs at $L \ll L^{\star}$. 
Extensive extragalactic surveys, e.g., SDSS \citep{blanton01,bell03} and CANDELS \citep{grogin11,koekemoer11}, and massively parallel hydrodynamical simulations \citep[e.g.,][]{vogelsberger14,vogelsberger20} have verified that the total populations of observed and simulated galaxies largely follow the above Schechter function at various redshifts. 
However, recent deep and wide galaxy surveys have shown that the Schechter function fails to reproduce the observed LFs at the bright ends \citep[e.g.,][]{parnovsky06,harikane22}. 
Additionally, \citet{hayashi20} demonstrated a prominent excess in the observed LFs of the ELGs compared to the Schechter function by selecting rare bright objects, leveraging the wide-field observation capabilities of the Subaru HSC. 
Recently, \citet{favole24} measured the LFs of ELGs, including the $\oii$-emitting galaxies, in the local Universe and found that the LFs of ELGs at $z<0.22$ are well described by the Saunders function \citep{saunders90}. 

We measure the LFs of our $\oii$ NB816 and NB921 samples to obtain strict constraints on the HOD parameters by comparing them to HOD-model predicted LFs (see Section~\ref{subsubsec:lf_from_hod}). 
The LFs of the $\oii$-emitting galaxies with $\log_{10}(L) \pm 0.5\Delta(\log_{10}(L))$, $\Phi(\log_{10}(L))$, are evaluated using the $1/V_{\rm max}$ method \citep{schmidt68} as follows:
\begin{equation}
\Phi(\log_{10}(L)) \Delta(\log_{10}(L)) = \sum_{i} \frac{P_{i}(\log_{10}(L))}{V_{\rm max} f_{\rm comp}},
\label{eq:lf_elg}
\end{equation}
where $V_{\rm max}$ corresponds to the survey volume, $f_{\rm comp}$ represents the completeness of ELGs that takes into account both selection and detection completeness, and $P_{i}(\log_{10}(L))$ is an expected intrinsic luminosity for each ELG \citep[for more details, refer to][]{hayashi18,hayashi20}. 
It is noted that the $1/V_{\rm max}$ method has been widely adopted and extensively tested in recent observational studies to robustly derive galaxy luminosity functions and stellar mass functions \citep[refer to][for more details]{takeuchi00,weigel16}. 

The observed line luminosities, $L_{\rm obs}$, are inevitably affected by dust attenuation and must be corrected to estimate the intrinsic luminosities of ELGs, $L_{\rm int}$. 
We adopt the dust-correction methods presented by \citet{cardelli89}, \citet{dominguez13}, and \citet{hayashi20} for our observed $\oii$ line luminosities as follows. 
The colour excess of the broad-band photometries, $E(B-V)$, can be related to that of the Balmer series H$\alpha$ and H$\beta$, $E({\rm H}\beta - {\rm H}\alpha)$, using the Balmer decrement \citep[see][]{momcheva13} as:
\begin{eqnarray}
E(B-V) & = & \frac{E({\rm H}\beta - {\rm H}\alpha)}{k(\lambda_{{\rm H}\beta}) - k(\lambda_{{\rm H}\alpha})} \nonumber \\
           & = & \frac{2.5}{k(\lambda_{{\rm H}\beta}) - k(\lambda_{{\rm H}\alpha})} \log_{10}\left(\frac{{({\rm H}\alpha/{\rm H}\beta)}_{\rm obs}}{{({\rm H}\alpha/{\rm H}\beta)}_{\rm int}} \right), 
\label{eq:colour_excess}
\end{eqnarray}
where ${({\rm H}\alpha/{\rm H}\beta)}_{\rm obs}$ and ${({\rm H}\alpha/{\rm H}\beta)}_{\rm int}$ represent the observed and intrinsic Balmer decrement, defined by the line flux ratios between H$\alpha$ and H$\beta$ lines, and $k(\lambda)$ is a dust attenuation curve at a wavelength $\lambda$. 
We use the dust reddening curve presented by \citet{calzetti00} for evaluating $k(\lambda)$. 
The observed Balmer decrement ${({\rm H}\alpha/{\rm H}\beta)}_{\rm obs}$ is evaluated as a function of stellar masses and line luminosities using the MPA-JHU release of spectrum measurements analysed for the SDSS Data Release 7 (DR7)\footnote{https://wwwmpa.mpa-garching.mpg.de/SDSS/DR7/} \citep[][see Appendix~2 of \citet{hayashi18} for more details about the observed Balmer decrement]{kauffmann03,salim07,abazajian09}. 
In contrast, the intrinsic Balmer decrement ${({\rm H}\alpha/{\rm H}\beta)}_{\rm int}$ is set to $2.86$, derived from Case B recombination given in \citet{osterbrock89}, assuming conditions of electron temperature $T_{\rm e} = 10^{4}$ K and electron density $n_{\rm e} = 10^{2}$ cm$^{-3}$. 
Using the colour excesses of each ELG and the reddening curve at the line wavelengths, the intrinsic line luminosities $L_{\rm int}(\lambda)$ can be estimated as:
\begin{eqnarray}
L_{\rm int}(\lambda) & = & L_{\rm obs}(\lambda) 10^{0.4 k(\lambda) E(B-V)} \nonumber \\
                               & = & L_{\rm obs}(\lambda) \frac{{({\rm H}\alpha/{\rm H}\beta)}_{\rm obs}}{{({\rm H}\alpha/{\rm H}\beta)}_{\rm int}}10^{\frac{k(\lambda)}{k(\lambda_{{\rm H}\beta}) - k(\lambda_{{\rm H}\alpha})}}, 
\label{eq:l_int}
\end{eqnarray}
where $L_{\rm obs}(\lambda)$ denotes the observed line luminosities of each ELG. 

We perform the JK resampling to evaluate the covariance matrices of LFs of $\oii$ NB816 and NB921. 
The settings of the JK resampling, i.e., the number of divisions into subregions, the shapes of each subregion, and the definition of the covariance matrix, are identical to those used for the ACFs given in Section~\ref{subsection:acf_and_cov}. 
The covariance matrices of the LFs are used to constrain the HOD-model parameters in Equation~\ref{eq:chi_lf}. 

It is worth noting that the LFs can be significantly biased if the survey volumes include notably over-dense regions \citep{takeuchi00}. 
To ensure that our LFs estimated by the 1/$V_{\rm max}$ method are not significantly affected by such large density fluctuations, we evaluate the number density distribution of ELGs in each JK subregion. 
The highest density subregion exhibited a density excess of $2.45\sigma$ for NB816 and $2.76\sigma$ for NB921 relative to the mean ELG number density. 
These results indicate that significant density fluctuations are absent in both redshift slices, ensuring that the 1/$V_{\rm max}$ method remains robust against potential biases. 

\section{A Theoretical Framework of our HOD model} \label{sec:hod_model} 
Our HOD model is built upon the framework proposed by \citet{geach12}, which includes two components for central ELG occupation: Gaussian and error function stochastic occupations, along with Poisson distributions for satellite ELGs. 
We revisit this model by incorporating observational evidence \citep[e.g.,][]{khostovan19,ishikawa20} and insights from other HOD models \citep[e.g.,][]{leauthaud12,cowley19,avila20}. 

Observational studies have revealed a tight correlation between host halo mass and the line luminosities of ELGs \citep{khostovan18,khostovan19,herrero23}, which is approximated by a double power-law model. 
The most significant feature of our new HOD model is the inclusion of the halo mass--line luminosity relation, allowing us to constrain the HOD parameters using observed LFs. 
Unlike the conventional HOD model that largely rely on galaxy abundance constraints, our model establishes a direct link between ELGs and their host haloes by constraining their occupation through differential baryonic number densities. 

The origin of the tight correlation between halo masses and line luminosities of ELGs is inferred from several well-known relationships. These include the main sequence of star-forming galaxies \citep[MS;][]{daddi07,lara10,renzini15}, the stellar-to-halo mass relation \citep[SHMR;][]{behroozi10,behroozi13,leauthaud11,leauthaud12}, and the star-formation rate (SFR)--line luminosity relation \citep[e.g.,][]{kennicutt12,saito20}. 
The MS represents a tight correlation between SFRs and stellar masses of star-forming galaxies, and both the MS and the SFR--line luminosity relation can be well approximated by single power-law models. 
On the other hand, the SHMR describes a correlation between stellar masses and halo masses with a double power-law model. 
Therefore, properties such as the double power-law feature and the transition scale of halo mass of the halo mass--line luminosity relationship can be largely influenced by the SHMR. 

\subsection{Central ELG occupation} \label{subsec:central_occupation}
While several HOD models incorporate the SHMR to describe the occupation of central galaxies \citep[e.g.,][]{leauthaud12,coupon15,cowley19}, our model takes a different approach by directly implementing the relationship between dark halo mass and line luminosity of ELGs. 
By doing so, we create a direct connection between the observable baryonic properties of ELGs and their spatial distributions within dark haloes. 

The central occupation number for a given halo mass $\Mh$ and luminosity threshold $\Lth$, denoted as $\Nc(\Mh |> \Lth)$, is defined as follows:
\begin{eqnarray}
\Nc(\Mh |> \Lth) = F_{\rm Gauss} \exp \left[-\frac{\log_{10}\left({f_{\rm LHMR} \left( \Mh \right) / L_{\rm Gauss}}\right)^{2}}{\left(\sigma_{{\rm log}M_{\rm h}} \right)^{2}}\right] \nonumber \nonumber \\
+ \frac{{F_{\rm erf}}}{2} \left[ 1 - {\rm erf}\left( \frac{\log_{10}{\left( \Lth / f_{\rm LHMR} \left( \Mh \right) \right)}}{\sigmalogM}\right) \right] \phi_{d}\left(\Mh \right). 
\label{eq:Nc}
\end{eqnarray}
The first term on the right-hand side represents the Gaussian distribution of central ELGs in less-massive haloes, which has been widely used to describe ELG occupation at the low-mass end \citep[e.g.,][]{geach12,avila20}. 
The second term describes the central ELG occupation in intermediate-to-massive haloes ($\Mh \gtrsim 10^{12},\Msun$) using the error function, as adopted by the standard \textit{vanilla} HOD model \citep[cf.,][]{zheng05,zheng07}. 
This combination of Gaussian and error function terms was explicitly shown by \citet{geach12} to successfully reproduce ELG occupation distributions derived from galaxy catalogues generated using the GALFORM semi-analytic model \citep{cole00}. 
The parameters $F_{\rm Gauss}$ and $F_{\rm erf}$ control the amplitudes of the Gaussian and the error functional central occupations, respectively. The widths of the Gaussian term and the transition scale of the error function are determined by $\sigma_{{\rm log}M_{\rm h}}$ and $\sigmalogM$, respectively. 
The parameter $L_{\rm Gauss}$ represents the line luminosity corresponding to the peak halo mass for the Gaussian occupation of central ELGs. 

In equation~\ref{eq:Nc}, we introduce two original functions, $f_{\rm LHMR}(\Mh)$ and $\phi_{\rm d}\left(\Mh \right)$. 
The function $f_{\rm LHMR}(\Mh)$ represents the luminosity-to-halo mass relation (hereafter, LHMR), which connects the luminosities of ELGs to the typical halo masses of their host haloes. 
This relationship is expressed by a double power-law model as follows:
\begin{eqnarray}
f_{\rm LHMR}(\Mh) &=& L_{\rm ELG}(\Mh) \nonumber \\
\label{eq:lhmr} &=& \frac{\Lzero}{2}\left[ \left(\frac{\Mh}{\Mtrans} \right)^{\beta} + \left( \frac{\Mh}{\Mtrans} \right)^{\gamma} \right], 
\end{eqnarray}
where $\beta$ and $\gamma$ are the power-law slopes for massive and less-massive ends of this relation, respectively. 
$\Mtrans$ represents the transition mass scale where two power laws intersect, and $\Lzero$ denotes the line luminosity around $\Mh \sim \Mtrans$. 
As mentioned above, the double power-law feature of the LHMR largely originates from the SHMR, and the halo mass parameter $\Mtrans$ in equation~\ref{eq:lhmr} is closely related to the pivot halo mass of the SHMR, referred to as $\Mpivot$. 
Both observational and theoretical studies have revealed that the pivot halo masses of the SHMR remain nearly constant around $\Mh \sim 10^{12} h^{-1}\Msun$ across a wide range of redshifts \citep[e.g.,][]{ishikawa16,ishikawa17,ishikawa20,behroozi19,legrand19}. 
From observational results \citep{khostovan19,herrero23}, it has been established that both power law slopes, $\beta$ and $\gamma$, are positive with the constraint, $0<\beta<\gamma$. 

The other original function, $\phi_{\rm d}\left(\Mh \right)$, controls the amplitude of central ELGs at the massive end and is described as follows:
\begin{equation}
\phi_{\rm d}\left(\Mh \right) = \frac{1}{2} \left[ 1 + F_{\rm d} {\rm erf} \left( \frac{\log_{10}\left({M_{\rm d}/\Mh}\right)}{\sigma_{\rm d}} \right) \right],
 \label{eq:decay}
 \end{equation}
where the parameter $F_{\rm d}$ represents the amplitude of the decay in central ELG occupation at the massive end. 
The characteristic mass scale at which the central ELG occupation begins to decay is denoted by $M_{\rm d}$, and $\sigma_{\rm d}$ parameterises the extent of this decay, determining how sharply the population of central ELGs declines as halo mass increases. 
 
In summary, the central occupation in our HOD model consists of the Gaussian occupation for less-massive haloes and the error function with a suppression at the high-mass end. This model results in 12 free parameters in total that characterise the central ELG occupation. 

\subsection{Satellite ELG occupation} \label{subsec:satellite_occupation}
The satellite occupation in our HOD model, $\Ns(\Mh |> \Lth)$, for a given luminosity threshold $\Lth$, follows a power-law distribution with a sharp exponential cutoff, as follows:
\begin{equation}
\Ns(\Mh |> \Lth) = F_{\rm s} \left(\frac{\Mh}{M_{\rm sat}} \right)^{\alpha_{\rm sat}} \exp{\left(- \frac{M_{\rm cut}}{\Mh} \right)}, 
\label{eq:Ns}
\end{equation}
where $F_{\rm s}$ is the amplitude of the power law, approximately representing the expected number of satellite ELGs within dark haloes of mass $M_{\rm sat}$. 
The parameter $\alpha_{\rm sat}$ controls the slope of the power-law behaviour for satellite ELG occupation, while $M_{\rm cut}$ defines the exponential cutoff mass, which limits satellite ELG occupation at the less-massive end. 

In line with the stellar-mass limited HOD model \citep{leauthaud12,cowley19}, our luminosity-limited model follows a similar relationship for the satellite mass scale $M_{\rm sat}$:
\begin{equation}
\frac{M_{\rm sat}}{10^{12} h^{-1}\Msun} = B_{\rm sat} \left( \frac{f_{\rm LHMR}^{-1} \left( \Lth \right)}{10^{12} h^{-1}\Msun} \right)^{\beta_{\rm sat}}, 
\label{eq:Msat}
\end{equation}
where $f_{\rm LHMR}^{-1}$ is the inverse of the LHMR, providing the corresponding halo masses for given galaxy luminosities. 
The normalisation parameter $B_{\rm sat}$ sets the mass scale for $M_{\rm sat}$, and $\beta_{\rm sat}$ determines the power-law scaling for this relation. 

Although our model for satellite occupation includes five free parameters, 
the following empirical relation can be used for the exponentially cutoff mass scale parameter $M_{\rm cut}$:
\begin{equation}
\log_{10}(M_{\rm cut}) = 0.76 \log_{10}(M_{\rm sat}) + 2.3. 
\label{eq:Mcut}
\end{equation}
This fitting formula comes from a tight correlation between halo mass parameters for satellite galaxy occupation introduced by \citet{conroy06}, who derived this relation from mock galaxy catalogues generated by the abundance matching approach between halo maximum circular velocities $V_{\rm max}$ in $N$-body simulations at $z=0-5$ and observed galaxy luminosity functions. 
In our model, $M_{\rm cut}$ and $M_{\rm sat}$ correspond to the mass parameters in the {\it vanilla} HOD model, making this empirical relation a reasonable choice for determining $M_{\rm cut}$. 

\subsection{Total ELG occupation} \label{subsec:total_occupation}
\subsubsection{Contamination-free total ELG occupation} \label{subsubsec:total_occupation}
The total ELG occupation function is obtained by summing the contributions from both central and satellite galaxies. 
However, we must also take into account the contamination in ELG samples. 
Fainter ELG candidates are susceptible to fake detections due to uncertainties in their line luminosities, and narrow-band filters may capture emission lines from unexpected sources. 
For instance, \citet{okada16} reported that about $4\%$ of their H$\alpha$ emitting galaxy samples were contaminated by other ELG populations. 

To account for these contaminating sources, we introduce the ELG fraction function, $f_{\rm ELG}(\Mh)$, and express the total occupation of ELGs with a halo mass $\Mh$ and a luminosity threshold $\Lth$ as follows:
\begin{eqnarray}
\NtotELG (\Mh |> \Lth) &=& \left( \Nc (\Mh |> \Lth) + \Ns (\Mh |> \Lth) \right) f_{\rm ELG}(\Mh) \nonumber \\
                                    &=& \Ntot (\Mh |> \Lth) f_{\rm ELG}(\Mh), 
\label{eq:Ntot}
\end{eqnarray}
where $\Ntot$ represents the total galaxy occupation, including contaminating samples. 

\begin{figure}
\includegraphics[width=\columnwidth]{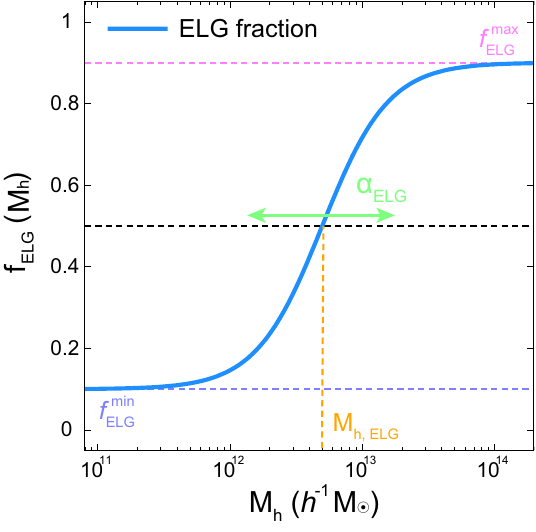}
\caption{The ELG fraction as a function of dark halo mass parameterized by equation~\ref{eq:f_elg}. This function is characterised by the standard sigmoid function described in equation~\ref{eq:sigmoid}. The lower and upper limit of the ELG fraction at the less-massive and massive end are controlled by $\fmin$ and $\fmax$ parameters, respectively. The median halo mass and the width of the transition scale are represented by $M_{h, {\rm ELG}}$ and $\alpha_{\rm ELG}$ parameters. In this plot, we assume each of four parameters as $\fmin = 0.10$, $\fmax = 0.90$, $M_{h, {\rm ELG}} = 5.0 \times 10^{12} h^{-1}\Msun$, and $\alpha_{\rm ELG} = 2.0$, respectively. }
\label{fig:f_elg}
\end{figure}

The function $f_{\rm ELG}(\Mh)$ is modelled as a sigmoid function, representing the {\it true} ELG fraction as a function of halo mass. 
We parameterise it as:
\begin{equation}
f_{\rm ELG}(\Mh) = \varsigma(\Mh) (\fmax - \fmin) + \fmin,
\label{eq:f_elg}
\end{equation}
where $\fmax$ is the maximum fraction of ELGs at the massive end, and $\fmin$ is the minimum fraction in less-massive haloes. 
These parameters are by definition bounded such that $0 \leq \fmin < \fmax \leq 1$. 
The function $\varsigma(\Mh)$ is a standard sigmoid function given by:
\begin{equation}
\varsigma(\Mh) =  \frac{{\rm tanh}\left( \alpha_{\rm ELG} \log_{10}(\Mh/M_{h, {\rm ELG}}) \right) + 1}{2}, 
\label{eq:sigmoid}
\end{equation}
where $M_{h, {\rm ELG}}$ is a halo mass at the transition scale of $\varsigma(\Mh)$, and $\alpha_{\rm ELG}$ controls the width of the transition around $M_{h, {\rm ELG}}$. 
Figure~\ref{fig:f_elg} illustrates the behaviour of $f_{\rm ELG}(\Mh)$ across different halo masses. 

We have introduced four additional parameters to describe the ELG fraction function. 
Thus, by assuming the empirical relation of equation~\ref{eq:Mcut}, our HOD model has a total of $20$ free parameters. 

\subsubsection{Fake ELG fraction} \label{subsubsec:fake_elg_fraction}
From the ELG fraction function (equation~\ref{eq:f_elg}), we can quantify the contamination of ELG samples by computing the total fake fraction $f_{\rm fake}$. 
The total fake fraction can be calculated as:
\begin{equation}
f_{\rm fake} = 1 - \frac{\int d\Mh \NtotELG(\Mh) \frac{dn}{d\Mh}}{\int d\Mh \Ntot(\Mh) \frac{dn}{d\Mh}},  
\label{eq:f_fake}
\end{equation}
where $\frac{dn}{d\Mh}$ represents the halo mass function. 

The ELG fraction function allows us to predict ACFs purely contributed by ELGs. 
Observed ACFs, however, are reduced in amplitude due to contaminants, making it necessary to correct for this effect \citep[e.g.,][]{hamana04,ishikawa17,okumura21}. 
Assuming that the contaminating samples are distributed homogeneously over the survey fields, we apply a correction factor to the HOD-predicted ACFs in the $i$-th angular bin, $\omega_{\rm HOD}(\theta_{i})$, using the total fake fraction $\ffake$. 
The contamination-corrected ACF is given by: 
\begin{equation}
\omega^{\rm corr}_{\rm HOD}(\theta_{i}) = (1 - f_{\rm fake})^{2} \omega_{\rm HOD}(\theta_{i}),
\label{eq:corr_acf}
\end{equation}
where $\omega^{\rm corr}_{\rm HOD}(\theta)$ is the corrected ACF that accounts for the reduction in amplitude caused by the contaminating fake ELG population. 

\subsubsection{Luminosity function from the HOD model} \label{subsubsec:lf_from_hod}
From the total ELG occupation function (equation~\ref{eq:Ntot}), we can predict the differential ELG LFs, $\Phi_{\rm HOD}(L_{1}, L_{2})$, which quantify the differential number density of ELGs with luminosities between $L_{1}$ and $L_{2}$, where $L_{1}$ < $L_{2}$. 
The LF is given by:
\begin{eqnarray}
\Phi_{\rm HOD}(L_{1}, L_{2}) &=& \frac{1}{L_{2} - L_{1}} \int d\Mh \frac{dn}{d\Mh} \nonumber \\
               &\times&  \left( \NtotELG(\Mh | > L_{1}) - \NtotELG(\Mh | > L_{2}) \right). \,\,\,\,\,\,\,\,
\label{eq:lf_HOD}
\end{eqnarray}
The above expression can be obtained by replacing the line luminosity threshold $L_{\rm th}$ in the first and second terms in the second line with $L_{1}$ and $L_{2}$, respectively. 
We can also obtain central and satellite ELG LFs by replacing the total occupation functions in equation~\ref{eq:lf_HOD} by those of central and satellite ELGs, respectively. 

The LF calculated by equation~\ref{eq:lf_HOD} entirely consists of ELGs; however, the observed LFs contain contributions from the contaminating sources. 
Therefore, we should use the total occupation function $\Ntot$ instead of the ELG occupation function $\NtotELG$ when comparing the HOD-predicted LFs with observed ones. 

\section{HOD-model analyses of $\oii$ emitting galaxies} \label{sec:hod}
In this section, we constrain the connection between $\oii$-emitting galaxies and dark matter haloes by analysing the observed ACFs of $\oii$ emitters from two approaches. 
First, we use the classical HOD model proposed by \citet{geach12}, a widely accepted model for describing the distribution of $\oii$ emitters within dark matter haloes. 
Second, we use our newly developed HOD model, which incorporates constraints from the LFs. 

\subsection{Common settings of the HOD analyses} \label{subsec:common_setting}
In this subsection, we present the settings and operations commonly used for both the Geach model and our new HOD model described in Section~\ref{sec:hod_model}. 
To obtain the ACFs from the HOD models, we use the public Python package, {\sc halomod}\footnote{https://halomod.readthedocs.io/en/latest/} \citep{murray21}. 
We have integrated our newly developed HOD model into this package. 
For HOD parameter estimation, we employ the Markov Chain Monte Carlo (MCMC) method using the publicly available Python library, {\sc emcee}\footnote{https://emcee.readthedocs.io/en/stable/} \citep{foreman13}. 
In our MCMC setup, we perform $600,000$ ($50,000$) MCMC steps for our HOD model (Geach HOD model) with $100$ walkers. 
The difference in the number of MCMC steps between models arises from the variation in the number of required steps to achieve convergence, which is influenced by the difference in the number of free parameters in the HOD models. 
We discard the first $30\%$ of steps as burn-in phases for each model. 
The redshift distributions for each population of $\oii$ emitters are assumed to be constant within the redshift ranges presented in Table~\ref{tab:elg}. 
Although this assumption is generally valid given the narrow redshift intervals ($\Delta z \sim 0.03$) of our samples, the effect of adopting  selection functions derived from narrow-band filter transmission curves was explicitly tested by \citet{okumura21} using the same $\oii$ emitter samples of NB816 and NB921. 
They found that the resulting amplitudes of the predicted ACFs differed by at most $\sim 7\%$, corresponding to only $\sim 4\%$ variation in the inferred fake fraction paremeter ($\tilde{f}_{\rm fake}$ in equation~\ref{eq:ng_corr}). 
Therefore, we conclude that our assumption of a constant redshift distribution has little impact on the inferred HOD parameters or on our main conclusions. 

We introduce a correction term to account for the integral constraint \citep[IC;][]{roche99} in the predicted ACFs from each of the HOD models. 
The observed ACFs tend to be underestimated at large angular scales due to the finite survey areas. 
In order to make a fair comparison between the model and actual observations, the ACFs should be corrected for the effect of the IC. 
The IC-corrected ACFs, denoted as $\omega^{\rm true}_{\rm HOD} (\theta)$, can be obtained using the ACFs corrected for the impact of fake-detected ELGs, $\omega_{\rm HOD}^{\rm corr} (\theta)$ (see equation~\ref{eq:corr_acf}), as follows:
\begin{equation}
\omega^{\rm true}_{\rm HOD} (\theta) = \omega_{\rm HOD}^{\rm corr} (\theta) - \frac{\sum_{i} \omega_{\rm HOD}^{\rm corr} (\theta_{i}) RR \left(\theta_{i}\right)}{\sum_{i} RR \left(\theta_{i}\right)}, 
\label{eq:acf_true}
\end{equation}
where $RR (\theta_{i})$ represents the normalised counts of random--random pairs within an angular separation of $\theta \pm \delta \theta$, weighted over four fields of the HSC Deep layers as:
\begin{equation}
RR (\theta_{i}) = \frac{\sum_{k=1}^{4} W_{k}^{2} \left( \theta_{i} \right) RR_{k} \left( \theta_{i} \right)}{\sum_{k=1}^{4} W_{k}^{2} \left( \theta_{i} \right) }, 
\label{eq:rr}
\end{equation}
where $RR_{k} \left( \theta_{i} \right)$ corresponds to the random-random pair of the $i$-th angular bin measured in the $k$-th subfield. 
The IC term is subtracted from the HOD-predicted ACF at each MCMC step because the impact of the IC term on $\omega^{\rm true}_{\rm HOD}$ varies depending upon the overall shapes of $\omega_{\rm HOD}^{\rm corr}$. 

We further apply a correction factor proposed by \citet{hartlap07} to the computed inverse covariance matrices of the ACFs and LFs (see Section~\ref{subsection:acf_and_cov} and \ref{subsec:lf}). 
This correction accounts for the finite number of JK resamplings and ensures that the inverse covariance matrix is unbiased. 
The corrected inverse covariance matrix, $\tilde{C}_{ij}^{-1}$ can be computed from the observed weighted average inverse covariance matrix, $C_{ij}^{-1}$, as follows:
\begin{equation}
\tilde{C}_{ij}^{-1} =  \frac{N_{\rm JK} - N_{\rm ACF} - 2}{N_{\rm JK} - 1} C_{ij}^{-1}, 
\label{eq:cov_hartlap}
\end{equation}
where $N_{\rm JK}$ and $N_{\rm ACF}$ represent the number of the JK resamplings and ACF bins, respectively. 
The same correction is also applied to the inverse covariance matrices of the LFs used by our HOD model. 
 
 Using the best-fitting HOD parameter set, we can extract several properties of the $\oii$ emitters and their host haloes as follows. 
The effective halo mass, which is a dark halo mass weighted by the number of ELGs, is given by
 \begin{equation}
 \Meff = \frac{\int d\Mh \Mh \NtotELG(\Mh) \frac{dn}{d\Mh}}{\int d\Mh \NtotELG(\Mh) \frac{dn}{d\Mh}}. 
 \label{eq:Meff}
 \end{equation}
The effective galaxy bias, a tracer bias calculated using the halo bias $\bh$ weighted by the number of ELGs, is given by
 \begin{equation}
 \bg = \frac{\int d\Mh \bh \NtotELG(\Mh) \frac{dn}{d\Mh}}{\int d\Mh \NtotELG(\Mh) \frac{dn}{d\Mh}}. 
  \label{eq:bg}
 \end{equation}
The satellite fraction represents the percentage of the satellite ELGs over total ELG abundance and is written as:
  \begin{equation}
 \fsat = \frac{\int d\Mh N_{\rm s, ELG}(\Mh) \frac{dn}{d\Mh}}{\int d\Mh \NtotELG(\Mh) \frac{dn}{d\Mh}}. 
  \label{eq:fsat}
 \end{equation}
 Finally, the mean number density of ELGs is derived by the halo mass function weighted by the total ELG occupation function as:
 \begin{equation}
 \ngal = \int d\Mh \NtotELG(\Mh) \frac{dn}{d\Mh}. 
  \label{eq:ngal}
 \end{equation}
 
We utilise the following analytical halo models when we predict ACFs and calculate the above deduced parameters from the HOD models. 
The halo mass function follows the analytical model proposed by \citet{tinker08}, the halo bias is a large-scale halo bias model of \citet{tinker10} with the scale dependence model of \citet{tinker05}, the halo concentration is assumed to follow the model of \citet{duffy08}, both the halo radial profile and distributions of satellite ELGs within haloes can be described by the NFW profile \citep{navarro97}, the transfer function is computed using the {\sc CAMB} package \citep{lewis00}, and the non-linear power spectrum is generated using the revised {\sc Halofit} model \citep{smith03,takahashi12}. 
The minimum and maximum halo mass of the integrations regarding to the halo mass are $10^{8} h^{-1}\Msun$ and $10^{15} h^{-1}\Msun$, respectively. 

\subsection{HOD-model fitting with the Geach model} \label{subsec:HOD_geach}
\subsubsection{Overview and setup} 
\label{subsubsec:geach_model}
The HOD model to explain the halo occupations of ELGs was introduced by \citet{geach12} using the mock galaxy catalogues obtained from semi-analytical galaxy formation models. 
Initially, it was developed to describe H$\alpha$-emitting galaxies selected by the HiZELS survey \citep{geach08}, but it has since been extended to various ELG populations \citep[e.g.,][]{cochrane18,hong19,okumura21,gao22}. 

The central ELG occupation of the Geach HOD model, $N_{\rm c}^{\rm Geach} (\Mh)$, consists of two parts: a Gaussian component and an error function component. 
The parameterised central ELG occupation can be expressed as follows:
\begin{eqnarray}
N_{\rm c}^{\rm Geach} (\Mh) &=& F_{\rm c}^{B} \left(1 - F_{\rm c}^{A}\right) \exp \left[-\frac{\log_{10}\left({\Mh / M_{\rm c}}\right)^{2}}{2\left(\sigma_{\log{M}} \right)^{2}}\right] \nonumber \\
                                              &+& F_{\rm c}^{A} \left[ 1 + {\rm erf}\left( \frac{\log_{10}{\left(\Mh / M_{\rm c} \right)}}{\sigma_{\log{M}}}\right) \right], 
\label{eq:nc_geach}
\end{eqnarray}
where $M_{\rm c}$ is the typical halo mass, and $\sigma_{\log{M}}$ describes the range for the central ELG distribution.
For the satellite occupation, $N_{\rm s}^{\rm Geach} (\Mh)$, the functional form is: 
\begin{equation}
N_{\rm s}^{\rm Geach} (\Mh)  = F_{\rm s} \left[ 1 + {\rm erf}\left( \frac{\log_{10}(\Mh/\Mmin)}{\delta_{\log{M}}} \right) \right] \left( \frac{\Mh}{\Mmin}\right)^{\alpha}, 
\label{eq:ns_geach}
\end{equation}
where $\Mmin$ is the characteristic halo mass to possess $F_{\rm s}$ satellite ELGs. 
Note that this $\Mmin$ is not the same definition as in the {\it vanilla} HOD model. 
As in the case of \citet{okumura21}, we fix $\delta_{\log{M}} = 1.0$ in equation~\ref{eq:ns_geach}. 

The total ELG occupation function, $N_{\rm tot}^{\rm Geach}(\Mh)$, is then:
\begin{equation}
N_{\rm tot}^{\rm Geach}(\Mh) = N_{\rm c}^{\rm Geach}(\Mh) + N_{\rm s}^{\rm Geach}(\Mh).
\label{eq:ntot_geach}
\end{equation}
Following the methodology outlined by \citet{okumura21}, we introduce the fake fraction parameter, $\tilde{f}_{\rm fake}$. 
This additional HOD parameter is varied to account for the total contamination ratio of the sample. 
It corrects the HOD-predicted ACFs by substituting $f_{\rm fake}$ with $\tilde{f}_{\rm fake}$ in equation~\ref{eq:corr_acf}. 
The number density of observed ELGs, $n_{\rm g}^{\rm obs}$, is also corrected for the effect of contaminations, and the corrected ELG abundance, $n_{\rm g}^{\rm corr}$, is expressed as:
\begin{equation}
n_{\rm g}^{\rm corr} = (1 - \tilde{f}_{\rm fake}) n_{\rm g}^{\rm obs}. 
\label{eq:ng_corr}
\end{equation}
In total, we have eight HOD-free parameters in the HOD analyses using the Geach model. 
Detailed descriptions can be found in \citet{geach12} and \citet{okumura21}. 

\begin{figure*}
\includegraphics[width=\columnwidth*2]{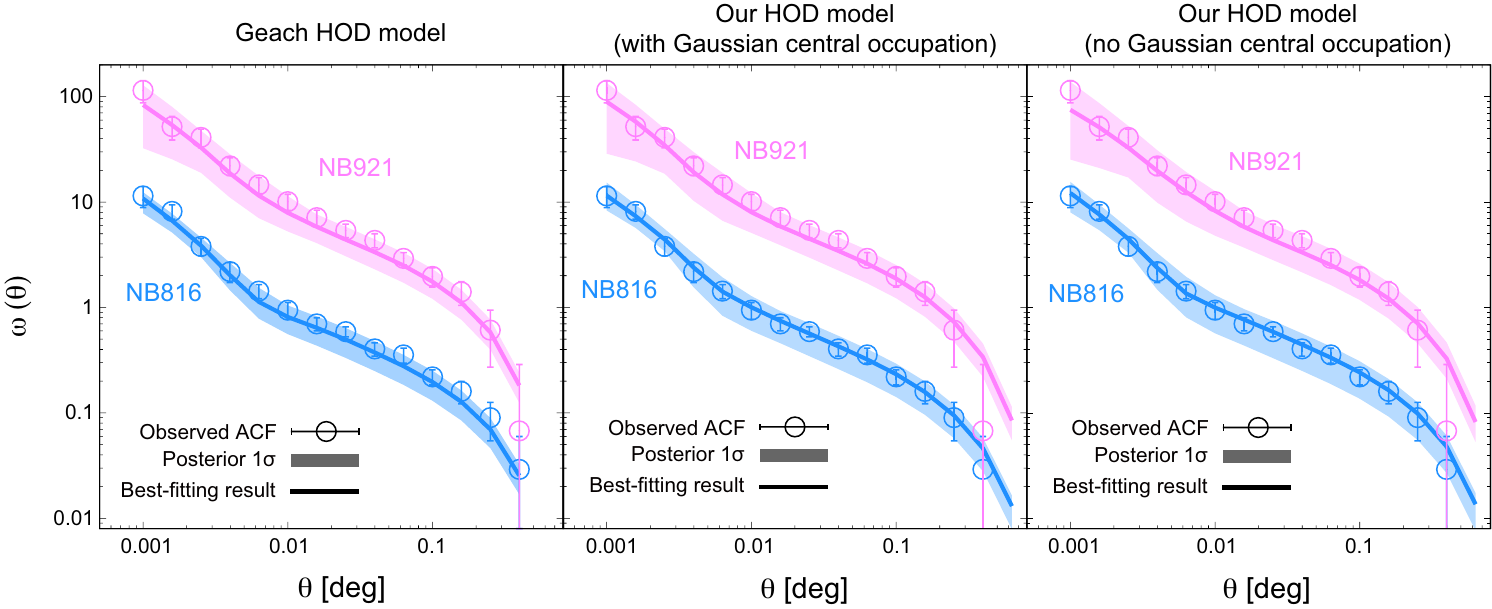}
\caption{The HOD-model fitting results for observed ACFs at $z=1.193$ (NB816; blue) and $z=1.471$ (NB921; red) predicted by the HOD model proposed by \citet[][left panel]{geach12} and this study (middle and right panels). Circles with error bars represent the observed ACFs of $\oii$-emitting galaxies. The solid lines depict the best-fitting results obtained through the entire MCMC fitting process, and the shaded regions indicate the ranges of ACFs corresponding to MCMC steps with $\chi^{2}$ values within the $1\sigma$ confidence level. The observed and HOD-predicted ACFs of NB921 are shifted upward by 1.0 dex for clarity in the plots.}
\label{fig:acf_hod}
\end{figure*}

Similar to the methodology used for our HOD model (see Section~\ref{subsec:HOD_lf} and \ref{subsec:HOD_lf_wo_gaussian}), the maximum likelihood estimation method is employed to constrain the HOD parameters. 
However, since this model is not linked to ELG line luminosities, we cannot constrain the ELG abundance through the LFs. 
Instead, we use the ELG number density constraint, and the contribution to $\chi^{2}$ from the ELG number density, $\chi_{n_{g}}^{2}$, can be written as follows:
\begin{equation}
\chi_{n_{g}}^{2} = \frac{\left( \log_{10}{n_{g}^{\rm corr}} - \log_{10}{\ngal} \right)^{2}}{\sigma_{\log{n_{g}}}^{2}}, 
\label{eq:chi_ng}
\end{equation}
where $\sigma_{\log{n_{g}}}$ is the uncertainty of the observed ELG number density. 
We adopt the relation, $\sigma_{\log{n_{g}}} = 0.03 | \log{n_{g}^{\rm obs}}|$, as in \cite{okumura21}. 
We fit the observed ACFs with those predicted by the Geach HOD model by replacing $\chi_{\rm LF}^{2}$ with $\chi_{n_{g}}^{2} $ in equation~\ref{eq:likelihood}. 

We employ a Gaussian prior on $\tilde{f}_{\rm fake}$ ($0.14 \pm 0.06$) and a flat prior for $\alpha$ in our HOD-fitting procedure, diverging from the approach of \citet{okumura21}, who introduced Gaussian priors for both $\tilde{f}_{\rm fake}$ and $\alpha$ parameters. 
The rationale behind not imposing a Gaussian prior on $\alpha$ stems from the absence of observational consensus on the formation efficiency for satellite ELGs in massive haloes. 
This difference in prior treatment can yield disparate results, despite making use of the identical ELG samples and the Geach HOD model, as in our study and those of \citet{okumura21}.

\subsubsection{HOD fitting results} \label{subsubsec:result_geach}
The left panel of Figure~\ref{fig:acf_hod} displays the results of the HOD-model fitting using the HOD model of \citet{geach12} on the observed $\oii$ NB816 and NB921 ACFs. 
The solid lines are the best-fitting ACFs throughout the overall MCMC fitting procedures. 
The shaded regions represent the maximum and minimum values of the ACFs among the MCMC steps where the chi-square, $\chi^{2}_{\rm step}$, satisfies, 
\begin{equation}
\chi^{2}_{\rm step} \le \chi^{2}_{\rm min} + \chi^{2}_{\rm dof},
\label{eq:chi2_dof}
\end{equation}
where $\chi^{2}_{\rm min}$ is the minimum chi-square value from the MCMC fitting, and $\chi^{2}_{\rm dof}$ corresponds to the $1\sigma$ confidence level determined by the degrees of freedom (dof). 
The dof is calculated as: 
\begin{equation}
{\rm dof} = N_{{\rm acf}} + N_{{\rm LF}} - k_{{\rm free}}, 
\label{eq:dof}
\end{equation}
where $N_{\rm acf}$ ($N_{\rm LF}$) represents the number of data points of the ACFs (LFs), and $k_{\rm free}$ denotes the number of free parameters of the HOD model. 
In the case of the HOD fitting using the Geach model, $N_{\rm LF} = 1$, corresponding to the one-point constraint from the number density of ELGs, and the number of free parameter in the HOD model is $ k_{{\rm free}} = 8$. 
The best-fitting and posterior mean HOD parameters are given in Table~\ref{tab:hod_geach}. 

As expected, the observed ACFs of $\oii$-emitting galaxies are well represented by the Geach HOD model. 
The reduced $\chi^{2}$ values solely from the fitting of ACFs are $\chi^{2}/{\rm dof} = 0.935$ for NB816 and $0.783$ for NB921, indicating that the observed ACFs are well reproduced by the model. 
However, it is important to emphasise that even if the ACFs are well reproduced by the HOD model, this does not necessarily imply that the galaxy occupation within dark halos is adequately captured by the HOD model \citep[see][for more details]{osato23}. 

\begin{table}
\caption{The best-fitting and posterior mean HOD and its deduced parameters using the HOD model of \citet{geach12}. }
\label{tab:hod_geach}
\begingroup
\renewcommand{\arraystretch}{1.18}
\begin{tabular}{lllll} \hline \hline
& \multicolumn{2}{c}{NB816 $(z_{\rm eff} = 1.193)$} & \multicolumn{2}{c}{NB921 $(z_{\rm eff} = 1.471)$} \\ \hline
 & Best fit & Posterior mean & Best fit & Posterior mean \\ \hline
$\log_{10}{(M_{\rm c})}^{a}$ & $11.50$ & $11.79^{+0.14}_{-0.15}$ & $11.48$ & $11.88^{+0.15}_{-0.16}$ \\
$\log_{10}{(M_{\rm min})}^{a}$ & $12.46$ & $12.80^{+0.37}_{-0.39}$ & $11.95$ & $11.04^{+0.34}_{-0.36}$ \\
$\sigma_{\log{M}}$ & $0.00$ & $0.07^{+0.06}_{-0.06}$ & $0.01$ & $0.09^{+0.06}_{-0.06}$ \\
$\alpha$ & $0.16$ & $0.25^{+0.19}_{-0.18}$ & $0.77$ & $0.50^{+0.33}_{-0.34}$ \\
$F_{\rm c}^{\rm A}$ & $0.19$ & $0.34^{+0.11}_{-0.11}$ & $0.12$ & $0.34^{+0.16}_{-0.12}$ \\
$F_{\rm c}^{\rm B}$ & $0.48$ & $0.39^{+0.35}_{-0.31}$ & $0.46$ & $0.41^{+0.35}_{-0.32}$ \\
$F_{\rm s}$ & $0.20$ & $0.38^{+0.29}_{-0.25}$ & $0.04$ & $0.47^{+0.34}_{-0.32}$ \\
$f_{\rm fake}$ & $0.01$ & $0.11^{+0.05}_{-0.05}$ & $0.03$ & $0.13^{+0.05}_{-0.05}$ \\ \hline \vspace{-0.1cm}
\multirow{2}{*}{$\chi^{2}/{\rm dof}$} & $7.52/8$ & $10.44/8$ & $6.18/8$ & $6.57/8$ \\ 
 & $= 0.94$ & $= 1.31$ & $= 0.77$ & $= 0.82$ \\
$\chi^{2}_{\rm ACF}$ & $7.43$ & $10.44$ & $6.16$ & $6.57$ \\
$\chi^{2}_{n_{g}}$ & $0.09$ & $0.01$ & $0.02$ & $0.00$ \\
$\log_{10}{(\bar{n}_{\rm ELG})}^{b}$ & $-2.23$ & $-2.25^{+0.07}_{-0.07}$ & $-2.42$ & $-2.44^{+0.07}_{-0.07}$ \\
$\log_{10}{(\Meff)}^{a}$ & $12.22$ & $12.42^{+0.07}_{-0.07}$ & $12.27$ & $12.45^{+0.08}_{-0.08}$ \\
$\bg$ & $1.42$ & $1.60^{+0.08}_{-0.08}$ & $1.67$ & $1.93^{+0.11}_{-0.11}$ \\
$\fsat$ & $0.25$ & $0.17^{+0.04}_{-0.04}$ & $0.20$ & $0.10^{+0.04}_{-0.04}$ \\ \hline
\multicolumn{5}{l}{\footnotesize$^a$ 
Halo mass parameters are in unit of $h^{-1}\Msun$ in logarithmic scale. 
} \\
\multicolumn{5}{l}{\footnotesize$^b$ 
ELG number densities are in unit of $(h^{-1}{\rm Mpc})^{-3}$ in logarithmic scale. 
} \\
\end{tabular}
\endgroup
\end{table}

\begin{figure}
\includegraphics[width=\columnwidth]{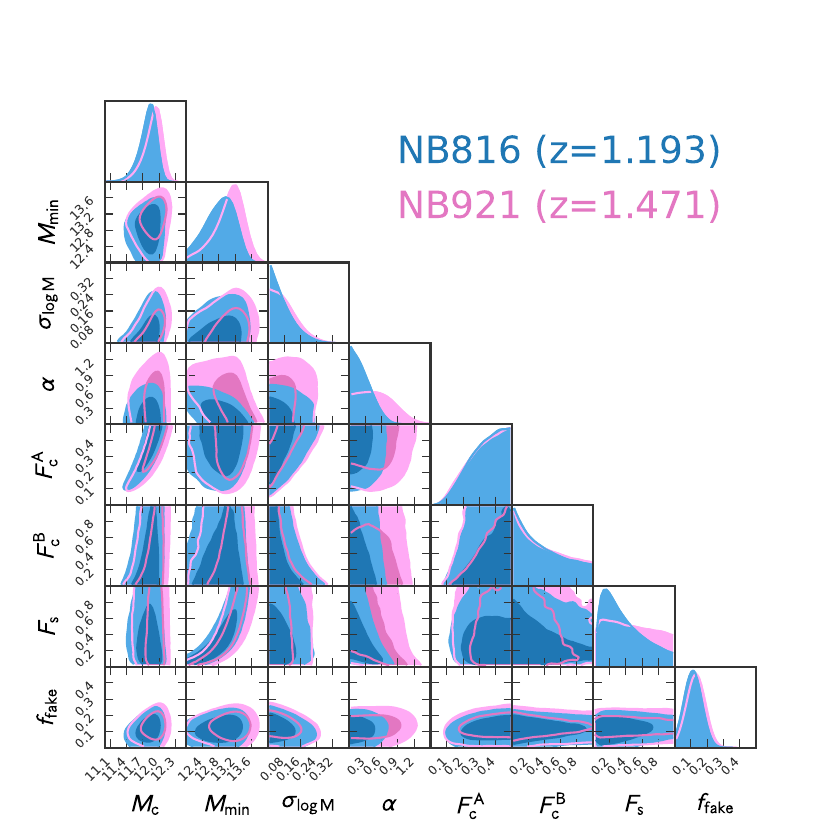}
\caption{The constrained HOD parameters of the HOD model proposed by \citet{geach12} obtained through the HOD-model fittings for $\oii$ emitting galaxies at $z=1.193$ (NB816; blue) and $z=1.471$ (NB921; red). The diagonal panels show 1D posterior probability distributions for each HOD parameter, whereas contours in other panels depict marginalised 2D posterior distributions. Dark and light-coloured regions of the contours correspond to the $68\%$ and $95\%$ confidence levels, respectively. Note that the HOD mass parameters, i.e., $M_{\rm c}$ and $M_{\rm min}$, are expressed in logarithmic scale units of $h^{-1}M_\odot$. }
\label{fig:contour_geach}
\end{figure}

The constrained HOD parameters from the ACF fitting using the Geach HOD model are presented in Figure~\ref{fig:contour_geach}. 
Similar to previous studies \citep[e.g.,][]{geach12,okumura21}, the normalisation parameters of central ($F_{\rm c}^{\rm A}$ and $F_{\rm c}^{\rm B}$) and satellite ($F_{\rm s}$) ELG occupations are not strongly constrained, but other parameters are successfully constrained. 
Notably, the parameter controlling satellite formation efficiency, $\alpha$, satisfies $\alpha<1$, suggesting that this model implies satellite ELGs are rarely formed within massive dark haloes at $z>1$. 

\begin{figure*}
\includegraphics[width=\columnwidth*2]{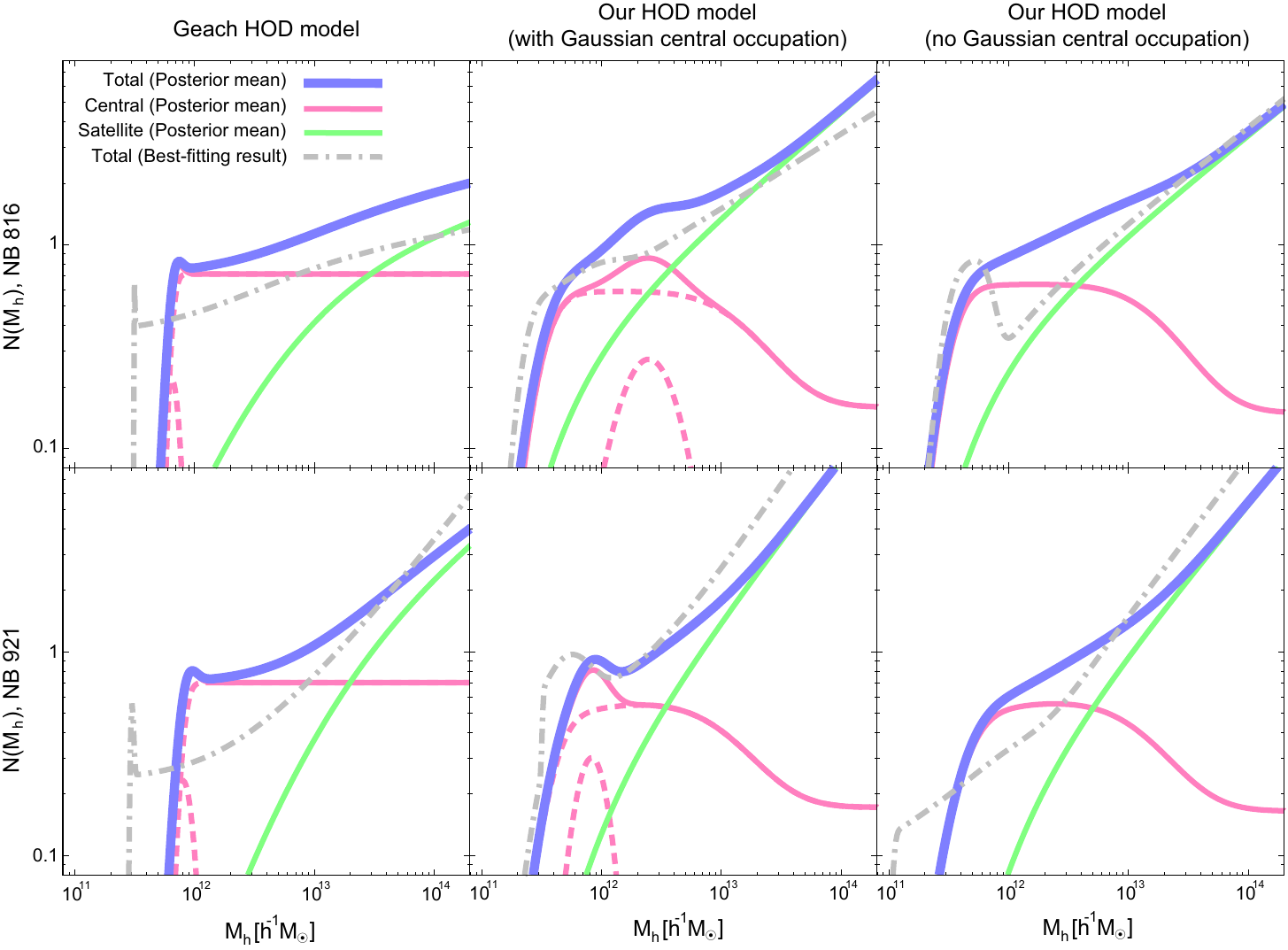}
\caption{The posterior mean occupation functions of $\oii$-emitting galaxies for NB816 (upper panels) and NB921 (lower panels) are shown as a function of host dark halo mass. The left, middle, and right panels display the results obtained using the HOD models proposed by \citet{geach12}, this study with Gaussian central occupation, and this study without Gaussian central occupation, respectively. In the left and middle panels, the total central occupations (solid red lines) consist of two components described by dashed red lines: the error functional occupations and the Gaussian occupations. The gray dash-dotted lines in each panel indicate the best-fitting total ELG occupations evaluated through the MCMC fitting procedures. }
\label{fig:hof}
\end{figure*}

The left panels of Figure~\ref{fig:hof} display the posterior mean and best-fitting occupation functions of $\oii$-emitting galaxies of NB816 (upper panel) and NB921(lower panel), evaluated using the Geach HOD model. 
In both populations, the posterior mean of the central ELG occupation rises steeply around $\Mh \sim 10^{12} h^{-1} \Msun$, with small contribution from the Gaussian component. 
Meanwhile, the best-fitting results of both populations also show a sharp increase at $\Mh \sim 3 \times 10^{11} h^{-1} \Msun$ with very narrow spreads, accompanied by a small contribution from the error function terms. 
These results imply that, adopting the Geach HOD model, the central ELGs instantaneously emerge in dark haloes that meet specific mass thresholds. 
However, this rapid formation process seems implausible unless there is a well-defined physical mechanism responsible for triggering the emission lines based on the halo mass. 

\begin{figure}
\includegraphics[width=\columnwidth]{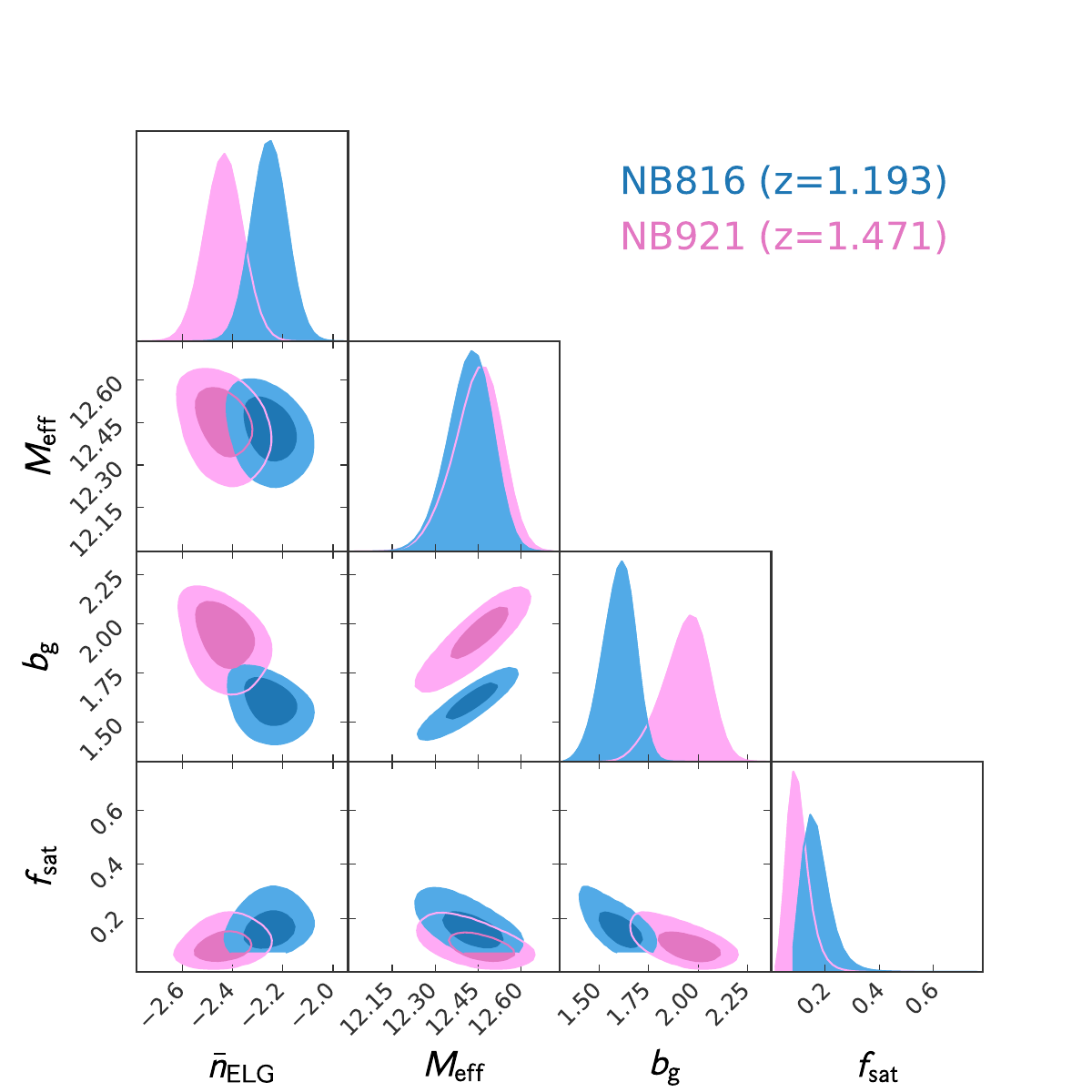}
\caption{The constrained deduced parameters obtained through the HOD-model fittings for $\oii$ emitting galaxies at $z=1.193$ (NB816; blue) and $z=1.471$ (NB921; red) using the model proposed by \citet{geach12}. Details of this figure are similar to Figure~\ref{fig:contour_geach}.}
\label{fig:deduced_geach}
\end{figure}
Figure~\ref{fig:deduced_geach} shows the parameters deduced from the HOD-model fitting using the Geach HOD model. 
The satellite fractions and the effective galaxy biases of $\oii$ emitting galaxies are evaluated as $\fsat = 0.17 \pm 0.04$ and $\bg = 1.60 \pm 0.08$ for NB816, and $\fsat = 0.10 \pm 0.04$ and $\bg = 1.93 \pm 0.11$ for NB921. 
Although our ELG samples and HOD model are identical to those in \cite{okumura21}, the constrained deduced parameters differ slightly. 
\citet{okumura21} reported $\fsat = 0.159^{+0.120}_{-0.047}$ ($0.159^{+0.109}_{-0.049}$) and $\bg = 1.700^{+0.084}_{-0.111}$ ($1.981^{+0.072}_{-0.068}$) for NB816 (NB921). 
This discrepancy could be attributed to differences in the treatment of priors. 
\citet{okumura21} applied a Gaussian prior on the $\alpha$ parameter ($\alpha = 1.00 \pm 0.20$), whereas we impose a flat prior on this parameter. 

We compare our inferred contamination fractions $f_{\rm fake}$ to studies of other emission-line galaxy populations to assess their plausibility. 
Although no published work has specifically measured contamination fractions for $\oii$ emitters, \citet{kusakabe18} reported a contamination level of $\sim 10 \pm 10\%$ for bright Ly$\alpha$ emitters at $z=2.14-2.22$, while \citet{okada16} found that their FastSound H$\alpha$ emitters $\sim 9\%$  in total. 
Our posterior mean values for the NB816 and NB921 filters, $0.11 \pm 0.05$ and $0.13 \pm 0.05$, respectively, therefore lie well within the range of contamination fractions identified in other ELG samples. 
This consistency suggests that a contamination level of around $10\%$ is not unexpectedly high for $\oii$-selected sources, thereby lending support to our treatment of contamination as a free parameter. 

\subsection{HOD-model fitting with our new model} \label{subsec:HOD_lf}
\subsubsection{Setup} \label{subsubsec:seup_lf_model}
Our HOD model includes $20$ free parameters in total, but two of them are fixed: the minimum and maximum values of the averaged true ELG fraction at the less-massive and massive ends, denoted as $\fmin$ and $\fmax$, respectively. 
These parameters are essential for describing the completeness and contamination levels of the ELG samples. 
We fix $\fmin=0.70$ and $\fmax=1.00$ for both the $\oii$ NB$816$ and NB$921$ populations to ensure consistent completeness thresholds across the analyses. 
Setting $\fmin=0.70$ reflects the goal of achieving $70\%$ completeness through the flux and magnitude limits, whereas setting $\fmax=1.00$ represents the completeness threshold. 

It is important to note, however, that completeness does not necessarily correspond to the true ELG fraction. 
For instance, observed faint ELGs may account for only $70\%$ of the total, but this does not imply that $30\%$ of the observed samples are contaminants. 
Instead, contaminants such as interloper galaxies, whose secondary emission lines are captured by the NB filter \citep{pullen16}, can be present regardless of their line luminosities. 
As a result, even bright ELG samples can include contaminants. 

Estimating the averaged true ELG fractions and interloper fractions solely from imaging data is challenging. 
Therefore, for the following analyses, we use the fixed values of $\fmin = 0.70$ and $\fmax = 1.00$. 
We have also conducted fitting analyses by varying $0.5\leq \fmin \leq 0.8$ and $0.9 \leq \fmax \leq 1.0$, and confirmed that this variation has little impact on the constraints of other parameters, with the final results remaining largely unchanged.

The remaining 18 parameters are constrained using the MCMC resampling technique, and we compare the derived ACFs and LFs from our model with observations through the maximum likelihood estimation. 
The likelihood is composed of two components of $\chi^{2}$ and is given by:
\begin{equation}
-2 \ln{\mathcal{L}} = \chi^{2}_{\rm ACF} + \chi^{2}_{\rm LF}, 
\label{eq:likelihood}
\end{equation}
where $\chi^{2}_{\rm ACF}$ and $\chi^{2}_{\rm LF}$ denotes the $\chi^{2}$ values obtained from the fittings on the ACF and LF components, respectively. 
The above types of $\chi^{2}$ can be computed using the $(i, j)$ element of the inverse covariance matrices of the ACFs and LFs, $\tilde{C}_{ij, {\rm ACF}}^{-1} $ and $\tilde{C}_{ij, {\rm LF}}^{-1} $, as follows:
\begin{equation}
\chi^{2}_{\rm ACF} = \sum_{i, j} \left( \omega_{\rm obs} \left( \theta_{i} \right) - \omega_{\rm HOD}^{\rm true} \left( \theta_{i} \right) \right) \tilde{C}_{ij, {\rm ACF}}^{-1} \left( \omega_{\rm obs} \left( \theta_{j} \right) - \omega_{\rm HOD}^{\rm true} \left( \theta_{j} \right) \right),
\label{eq:chi_acf}
\end{equation}
and 
\begin{equation}
\chi^{2}_{\rm LF} = \sum_{i, j} \left( \Phi_{\rm obs} \left( L_{i} \right) - \Phi_{\rm HOD} \left( L_{i} \right) \right) \tilde{C}_{ij, {\rm LF}}^{-1} \left( \Phi_{\rm obs} \left( L_{j} \right) - \Phi_{\rm HOD} \left( L_{j} \right) \right). 
\label{eq:chi_lf}
\end{equation}
During the MCMC fitting procedures, we assume flat priors for all the HOD parameters. 

In addition to the derived parameters shown in equation~\ref{eq:Meff}-\ref{eq:ngal}, we also introduce an additional deduced parameter, $M_{\rm min}^{\rm LF}$, representing the median halo mass of central ELGs. 
This mass parameter is defined similarly to the $\Mmin$ parameter in the {\it vanilla} HOD model. 
The $\Mmin$ parameter is widely used in clustering studies to discuss the formation and evolution of galaxies as a characteristic halo mass. 
Introducing the $M_{\rm min}^{\rm LF}$ parameter is highly beneficial for examining the relationship between ELGs and other galaxy populations, which are primarily derived from the {\it vanilla} HOD model. 
We define the $M_{\rm min}^{\rm LF}$ as follows:
\begin{equation}
\frac{N_{\rm c, ELG}(M_{\rm min}^{\rm LF})}{\max N_{\rm c, ELG}(\Mh)} = 0.5, 
\label{eq:Mmin_elg}
\end{equation}
where $\max N_{\rm c, ELG}(\Mh)$ represents the maximum expected number of central ELGs, with $\max N_{\rm c, ELG}(\Mh) \leq 1.0$ by definition. 
In our HOD model, two different halo masses can satisfy the above definition: one is the lower-mass halo where the central occupation is in the middle of increasing according to the error function, and the other is the higher-mass halo where the central occupation decreases following the decaying function $\phi_{\rm d}(\Mh)$. 
To meet the definition of the original $\Mmin$ parameter, we use the least massive halo as $M_{\rm min}^{\rm LF}$ when multiple masses satisfy equation~\ref{eq:Mmin_elg}. 

\subsubsection{HOD fitting results} \label{subsubsec:result_lf}
The middle panel of Figure~\ref{fig:acf_hod} shows the fitting results of observed ACFs with our HOD model. 
Since we fix two HOD parameters, $\fmin$ and $\fmax$, and $k_{\rm free}$ in equation~\ref{eq:dof} is $18$, 
the dof of NB816 and NB921 samples are ${\rm dof} = 15$ (NB816) and $14$ (NB921), respectively. Our HOD model that contains LF constraints successfully reproduces the observed ACFs of ELGs at the wide angular ranges. 
The minimum values of reduced $\chi^{2}$ are $\chi^{2}/{\rm dof} = 0.79$ (NB816) and $0.86$ (NB921), showing that we have reasonable fitting results using our HOD model. 

\begin{figure}
\includegraphics[width=\columnwidth]{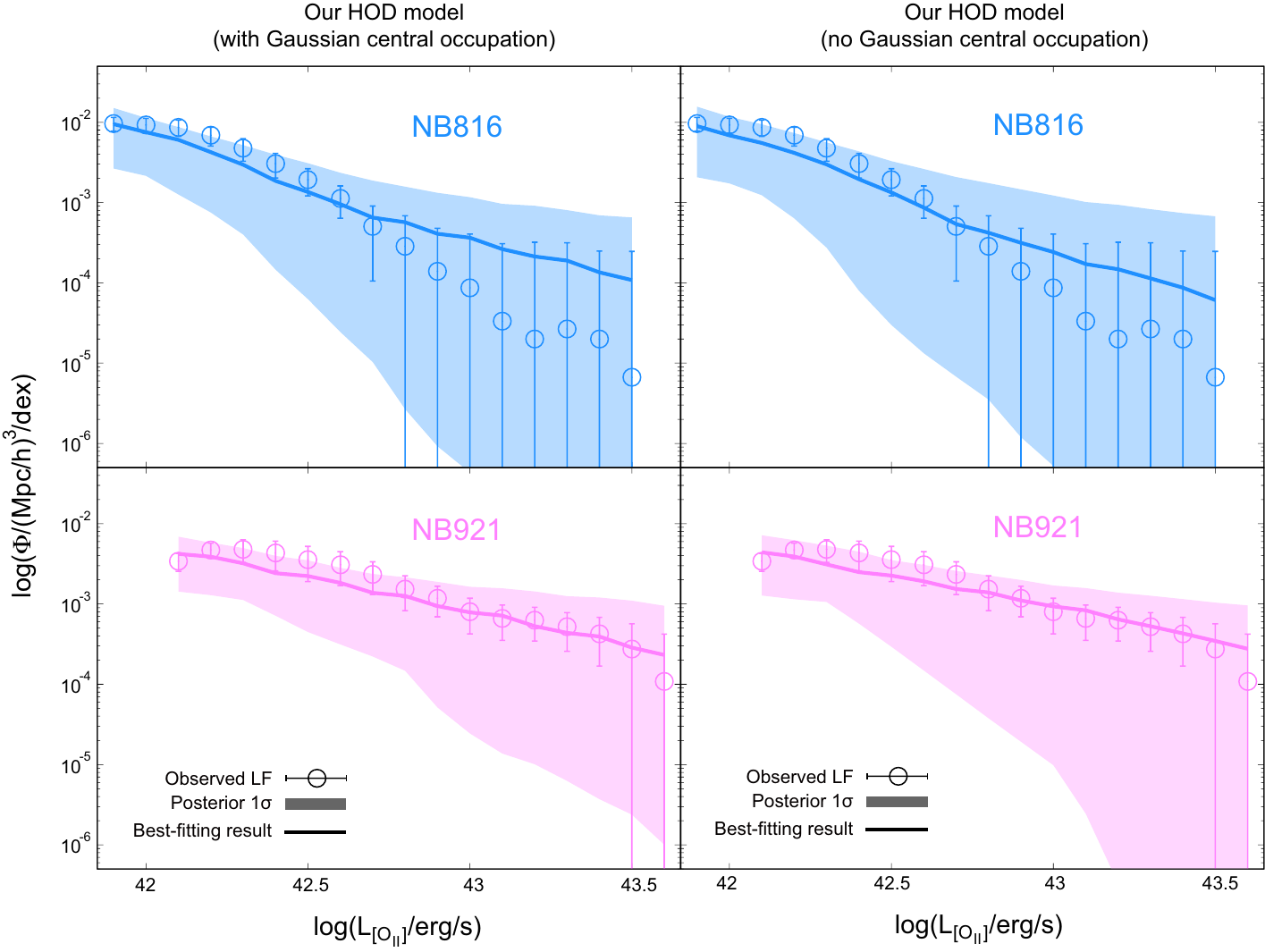}
\caption{The HOD-model fitting results for observed luminosity functions at $z=1.193$ (NB816; blue) and $z=1.471$ (NB921; red) using our HOD model. Left panels show the results using our HOD model with the Gaussian central occupation (Section~\ref{subsubsec:result_lf}), whereas right panels are results using our HOD model without the Gaussian central occupation (Section~\ref{subsec:HOD_lf_wo_gaussian}). Details of this figure are similar to Figure~\ref{fig:acf_hod}.}
\label{fig:lf_ishikawa}
\end{figure}

The fitting results of the LFs are shown in Figure~\ref{fig:lf_ishikawa}. 
Our HOD model is flexible since we do not assume any functional form of the LFs. The best-fitting results of the LFs are fully consistent with the observed LFs. 
Our observed LFs have relatively large uncertainties, especially at the bright ends, since we calculate LFs by varying the line luminosity ratio ${\rm H}\alpha/{\rm H}\beta$, which are calibrated by the local SDSS galaxies, within $1\sigma$ confidence intervals. 
These large uncertainties at the bright ends prevent from strict constraints on the HOD parameters from the LFs. 
However, joint analysis of clustering with the LF provides a unique constraint from the differential number densities about the line luminosities compared to the total number density constraint, and it helps to connect the ELG host haloes to baryonic properties of ELGs and unveil the evolutionary path of ELGs (See Section~\ref{sec:discussion}). 

Contamination from misidentified $\oii$ emitters is explicitly accounted for in our modelled LFs using a contamination-corrected occupation function (equation~\ref{eq:Ntot}). 
The inferred contamination fractions from our analysis are consistent with independent estimates obtained from spectroscopic observations \citep[e.g.,][]{okada16,kusakabe18}, suggesting that the expected biases in our results due to contamination are small. 

\begin{figure*}
\includegraphics[width=\textwidth]{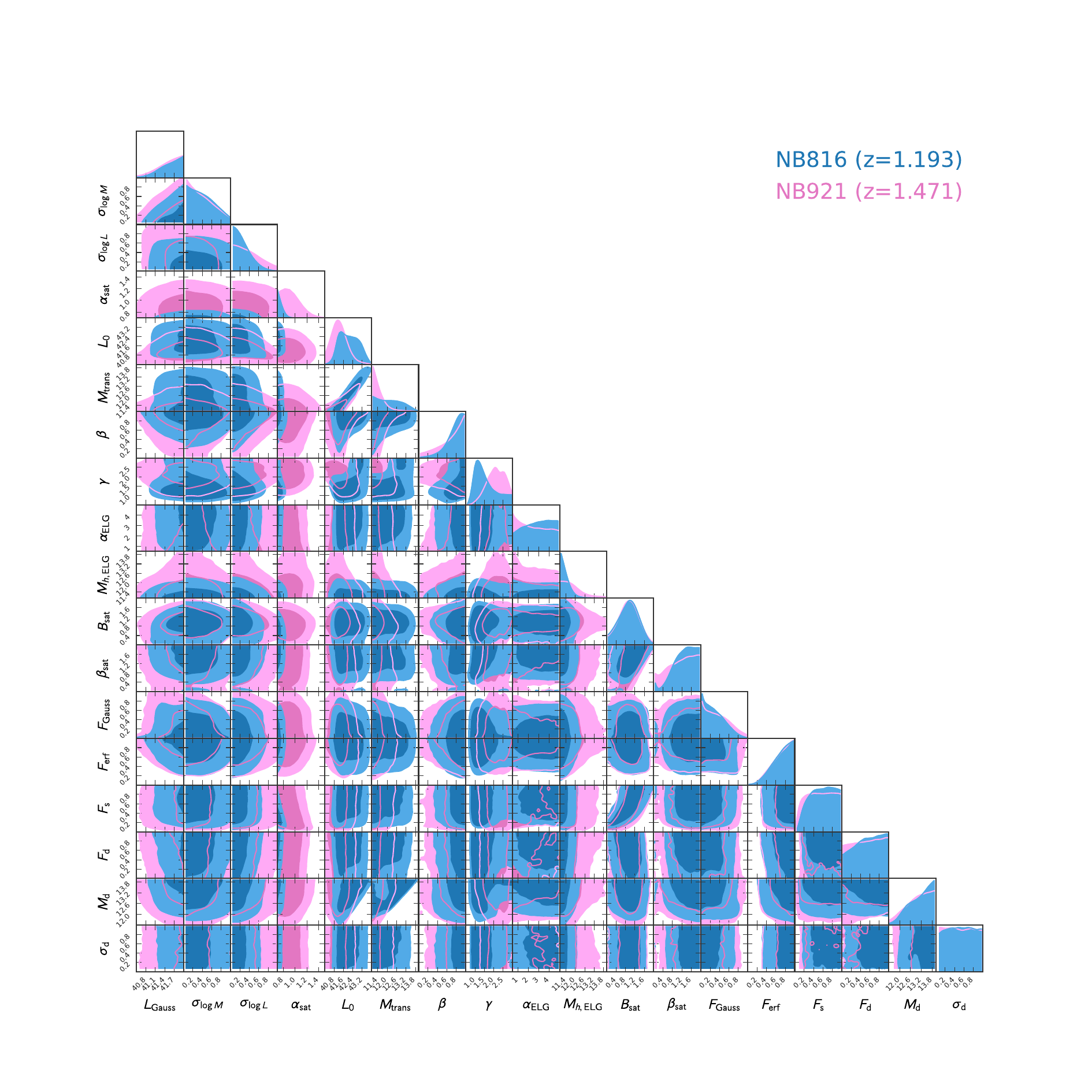}
\caption{Similar to Figure~\ref{fig:contour_geach}, but showing the constrained HOD parameters adopting our HOD model with the Gaussian central occupation. }
\label{fig:contour_ishikawa_lf}
\end{figure*}

The HOD parameters constrained by the MCMC fittings are presented in Figure~\ref{fig:contour_ishikawa_lf}. 
The amplitude of the Gaussian central occupation, $F_{\rm Gauss}$, is found to be small, constrained to $F_{\rm Gauss} < 0.4$ for both redshift slices. 
This indicates that the Gaussian central occupation is indeed negligible for the total central ELG occupation, consistent with previous HOD studies focused on $\oii$ emitters \citep[e.g.,][]{okumura21,osato23}.

The ELG occupations within dark haloes predicted by our HOD model are shown in the middle panels of Figure~\ref{fig:hof}. 
Unlike the results derived from the Geach HOD model, our HOD model predicts a smoother transition for central ELG occupations from lower-mass haloes to Milky-Way-sized haloes. 
However, similar to the results using the Geach HOD model, the contributions from the Gaussian central occupations are relatively small compared to those of the error functional central occupations. 
Consequently, the impact of the Gaussian term can be considered negligible for the total central occupations. 
In Section~\ref{subsec:HOD_lf_wo_gaussian}, we will further investigate whether our HOD model can still reproduce the observed ACFs and LFs by excluding the Gaussian central occupation. 
We will then compare the results obtained with and without the inclusion of the Gaussian central occupation to assess any significant differences. 

\subsubsection{Deduced parameters}
\begin{figure}
\includegraphics[width=\linewidth]{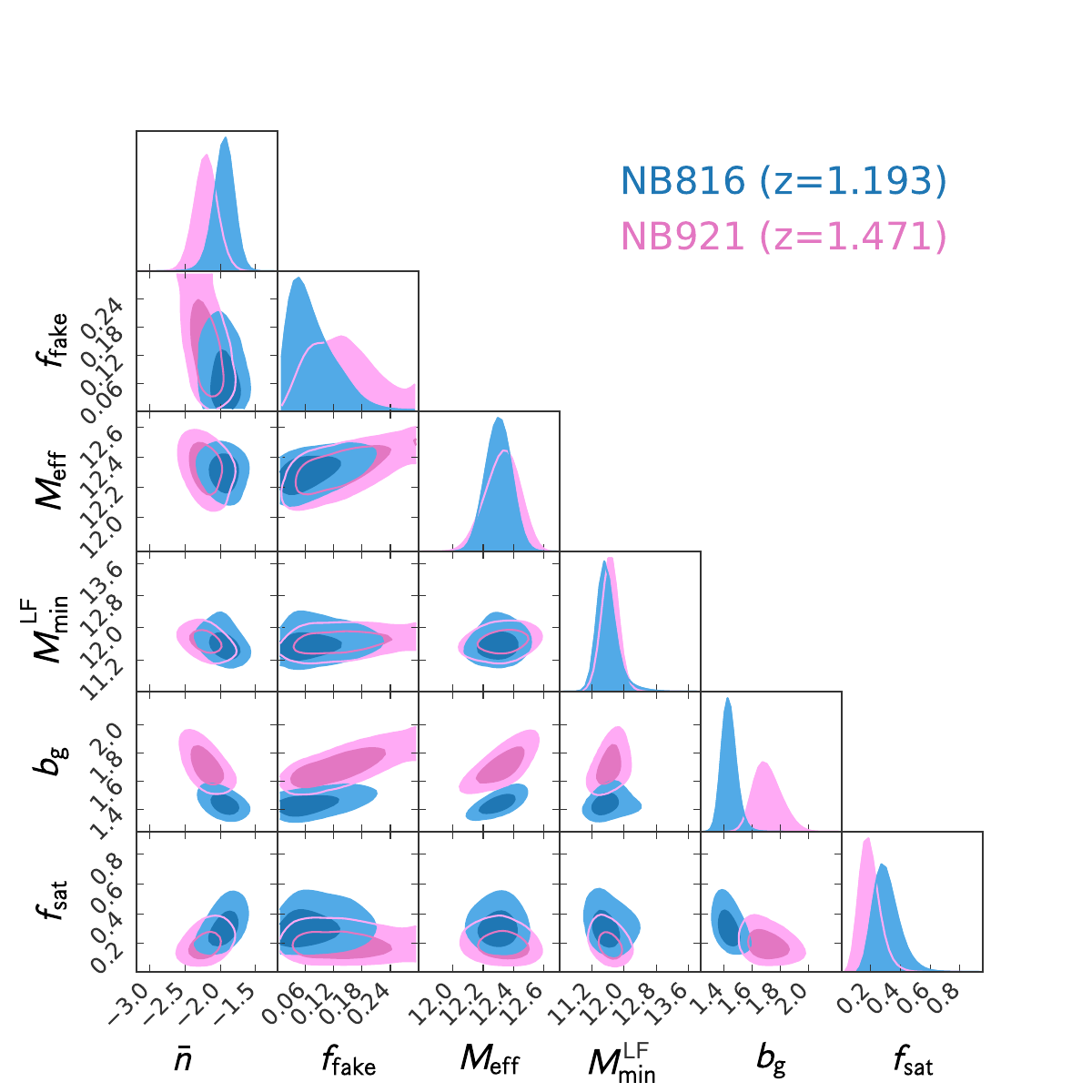}
\caption{Similar to Figure~\ref{fig:deduced_geach}, but showing the deduced HOD parameters adopting the HOD model proposed by this study with the central Gaussian occupation. }
\label{fig:contour_deduced_ishikawa_lf}
\end{figure}
Figure~\ref{fig:contour_deduced_ishikawa_lf} shows the constraints on the deduced parameters, $\ngal$, $f_{\rm fake}$, $\Meff$, $M_{\rm min}^{\rm LF}$, $\bg$, and $\fsat$, inferred from the constrained HOD parameters. 
The median halo masses of the ELG host haloes, $M_{\rm min}^{\rm LF}$, are evaluated as $\log_{10}(M_{\rm min}^{\rm LF}/h^{-1}\Msun) = 11.60^{+0.19}_{-0.20}$ for the NB816 sample and $11.66^{+0.18}_{-0.19}$ for the NB921 sample. 
These results are consistent with the characteristic halo masses of ELGs obtained by other extensive surveys \citep[e.g.,][]{avila20,gao22,rocher23a} and simulations \citep[e.g.,][]{rocher23b}, although the definitions of the characteristic halo masses in literature do not necessarily identical to $M_{\rm min}^{\rm LF}$. 

The satellite fractions, $\fsat$, of our $\oii$-emitting galaxies are estimated to be $\fsat = 0.31 \pm 0.08$ and $0.20 \pm 0.06$ for the NB816 and NB921 samples, respectively. 
Since our ELG samples are selected using the same line flux threshold, i.e., $f_{\rm th} = 3.0 \times 10^{-17}$ ${\rm [erg/s/cm^{2}]}$, the differences in satellite fractions purely reflect the redshift evolution of $\oii$-emitting galaxies. 
Therefore, the lower fraction of satellite ELGs at slightly higher redshifts is attributed to the lower abundance of massive haloes that can host multiple satellite ELGs. 
This trend has been observed in other HOD studies exploring the redshift evolution of satellite fractions as a function of stellar mass and absolute magnitude within the same galaxy populations \citep[e.g.,][]{zehavi11,coupon12,mccracken15,ishikawa17,ishikawa20,ishikawa21}. 
Our results indicate that ELGs also exhibit the same decreasing trend in satellite fraction with increasing redshift, similar to other galaxy populations. 

The effective galaxy bias, $\bg$, of our ELG samples show $\bg = 1.44 \pm 0.05$ (NB816) and $1.71 \pm 0.09$ (NB921). 
\citet{okumura21} calculated the bias parameters of ELGs using the same ELG catalogue of this study from the HOD analysis adopting Geach HOD model, and reported that $\bg = 1.70^{+0.08}_{-0.11}$ (NB816) and $1.98 \pm 0.07$ (NB921). 
Recently, \citet{rocher23a} conducted the HOD model analysis on ELGs selected by the One-Percent survey of the Dark Energy Spectroscopic Instrument \citep{desi22} and calculated the linear bias parameters by generating DESI mock ELG catalogues using the constrained HOD parameters. 
The resulting linear bias parameters for ELGs were $\bg = 1.33 \pm 0.03$ at $z=1.1$ and $1.45 \pm 0.03$ at $z=1.325$. 
By comparing these studies with our results, our bias parameters are larger than those from \citet{rocher23a}, but smaller than the results of \citet{okumura21}. 
The differences in these galaxy bias values may be attributed to several factors, such as differences in sample selection, redshift range, or the specific halo occupation model employed in each study. 

The fake fraction, $\ffake$, is evaluated as $\ffake = 0.08 \pm 0.05$ and $0.14 \pm 0.07$ for the NB816 and NB921 samples, respectively. 
\citet{okumura21} introduced the fake fraction parameter as a free parameter in the HOD-fitting procedure (see Section~\ref{subsubsec:geach_model}) and found that the fake fractions of $\oii$-emitting galaxies evaluated using the Geach HOD model are $\ffake = 0.140^{+0.048}_{-0.053}$ (NB816) and $0.104^{+0.043}_{-0.041}$ (NB921). 
\citet{hayashi20} estimated the contamination fraction of NB816 and NB921 $\oii$-emitting galaxies using spectroscopically observed samples and concluded that the contamination fractions are about $10 - 20 \%$. 
Consequently, the contamination fractions derived from our HOD model analyses and those from \citet{okumura21} are consistent with these spectroscopic estimates, indicating that our results are reasonable and in good agreement with previous studies. 

In addition to the ELG--LRG evolutionary connection detailed in the following discussion (see Section~\ref{subsubsection:satellite_fraction}), our inferred HOD parameters indicate that $\oii$ emitters at $z=1.193$ and $1.471$ reside in systematically different halo environments from more evolved galaxy populations. 
Specifically, we find lower effective halo masses and relatively higher satellite fractions ($\fsat \approx 0.2-0.4$) than those typically measured for red-sequence at the same redshift ranges, which often show $\fsat < 0.15$ \citep{coupon12,mccracken15,ishikawa21}. 
Such characteristics highlight the star-forming, younger nature of ELGs and underscore their importance for large-scale structure studies. 
For instance, accurate modelling of ELGs, which preferentially populate lower-mass haloes with relatively abundant satellites, will be essential in future galaxy surveys for obtaining unbiased cosmological constraints on the cosmic expansion history through baryon acoustic oscillation and redshift-space distortion measurements. 
Additionally, the higher satellite fractions inferred for ELGs suggest their potential utility for detailed environmental studies of galaxy groups and cluster outskirts, and may provide valuable observational constraints on theoretical models of galaxy quenching and environmental evolution. 

\subsection{HOD-model fitting with our model (without Gaussian central occupation)} \label{subsec:HOD_lf_wo_gaussian}
As seen in the previous section, we have found that contributions of the Gaussian component to the total central occupation is small compared to that of the error function components. 
In this section, we demonstrate the HOD-model fitting with our modified model, where the Gaussian central occupation is excluded from the total central occupation, and compare the results with those obtained in Section~\ref{subsec:HOD_lf}. 

\begin{figure*}
\includegraphics[width=\linewidth]{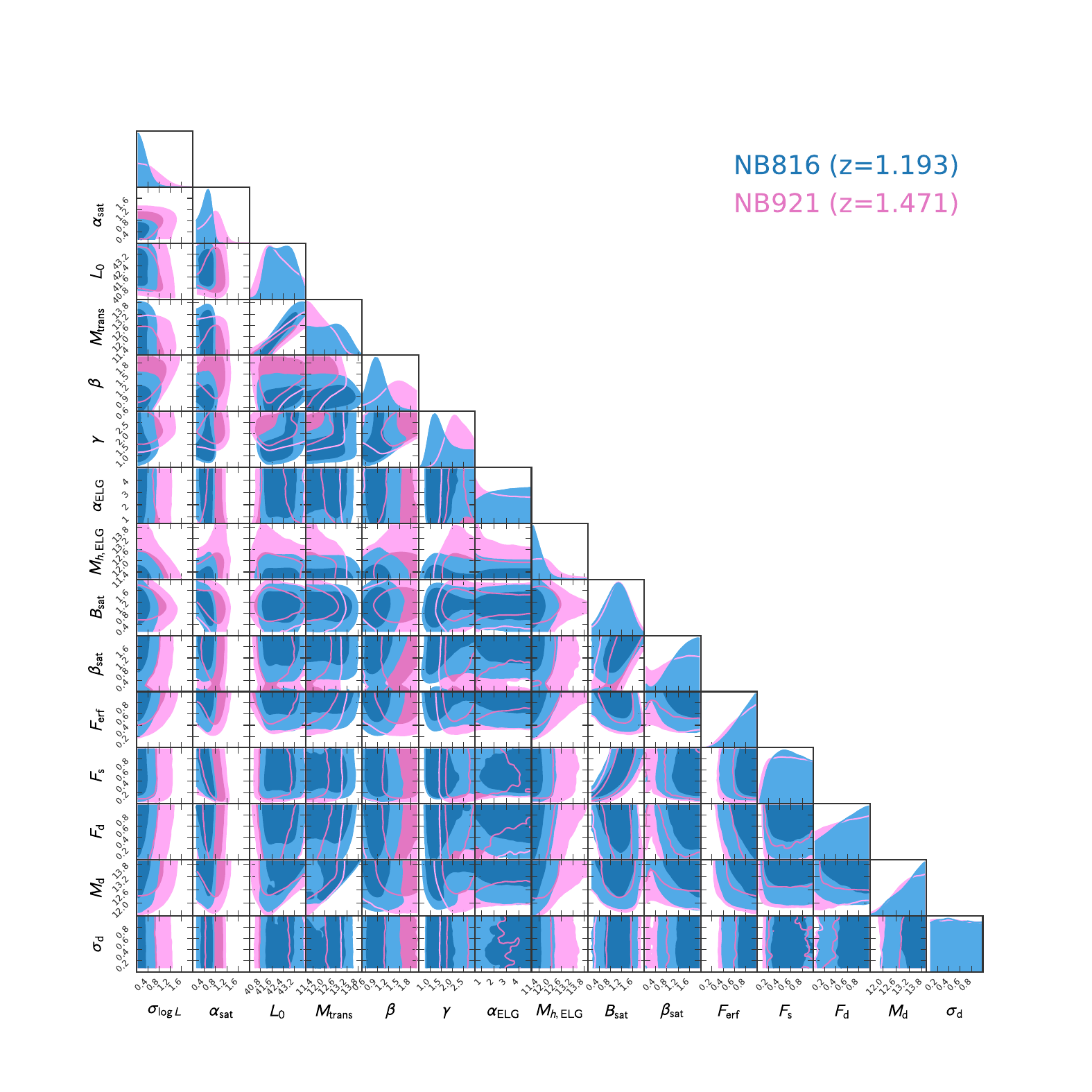}
\caption{Similar to Figure~\ref{fig:contour_geach}, but showing the constrained HOD parameters adopting our HOD model without the Gaussian central occupation by setting $F_{\rm Gauss} = 0.0$. In this plot, the HOD parameters that control the Gaussian central occupation, i.e., $F_{\rm Gauss}$, $L_{\rm Gauss}$, and $\sigmalogM$, are absent compared to Figure~\ref{fig:contour_ishikawa_lf}. }
\label{fig:contour_ishikawa_wo_gaussian}
\end{figure*}

We set $F_{\rm Gauss} = 0$ to exclude the Gaussian central occupation in Equation~\ref{eq:Nc} and perform the HOD-model analysis following the same procedure as in Section~\ref{subsec:HOD_lf}. 
This setting eliminates three free parameters related to the central occupation function: $F_{\rm Gauss}$, $L_{\rm Gauss}$, and $\sigma_{{\rm log}M_{\rm h}}$. 
The best-fitting ACFs and LFs obtained with this modified model are shown in the right panels of Figures~\ref{fig:acf_hod} and \ref{fig:lf_ishikawa}, respectively. 
The best-fitting and posterior mean HOD parameters, along with the corresponding values of the reduced $\chi^{2}$, are listed in Table~\ref{tab:hod_lf}. 
The parameter contours constrained by the MCMC fittings are shown in Figure~\ref{fig:contour_ishikawa_wo_gaussian}. 

Comparing the fitting results from this simplified model to those from our original model with the Gaussian central occupation, we find that although the $\chi^{2}$ values are sufficiently small for both models, indicating good fits, the best-fitting results of our model without the Gaussian central occupation achieve better fits for both populations. 
This suggests that while our original model allows for more flexible fitting compared to the simplified model, the improvement in reproducing the observational results is insufficient relative to the increase in the number of parameters. 
The halo occupation functions derived from this simplified model are presented in the right panels of Figure~\ref{fig:hof}. 
Our simplified model provides reasonable halo occupation functions compared to those obtained using the Geach HOD model (left panels of Figure~\ref{fig:hof}), particularly in terms of the smooth increase in less-massive central ELGs and satellite ELGs in massive dark haloes.

\begin{figure}
\includegraphics[width=\linewidth]{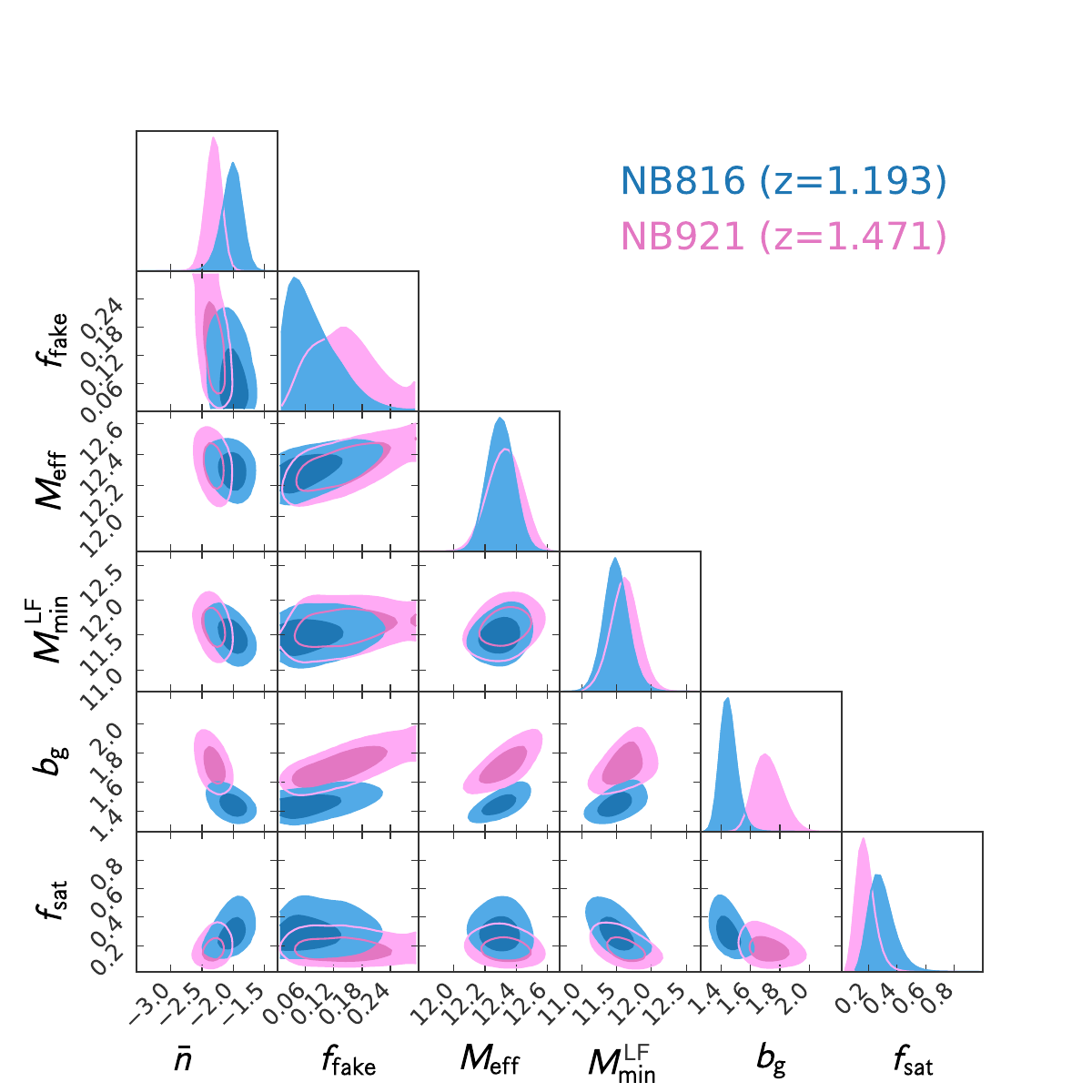}
\caption{Similar to Figure~\ref{fig:deduced_geach}, but showing the deduced HOD parameters adopting the HOD model proposed by this study without the central Gaussian occupation. }
\label{fig:contour_deduced_ishikawa_lf_wo_gauss}
\end{figure}

The deduced parameters are also summarised in Table~\ref{tab:hod_lf}, and their 1D marginalised posterior distributions, along with 2D confidence contours, are shown in Figure~\ref{fig:contour_deduced_ishikawa_lf_wo_gauss}. 
Compared to the results in Figure~\ref{fig:contour_deduced_ishikawa_lf}, the deduced parameters obtained from both models are largely consistent within the $1\sigma$ confidence levels. 
For instance, the characteristic dark halo mass for $\oii$ NB816 (NB921) evaluated by the original and simplified central HOD models are $\log_{10}{(M_{\rm min}^{\rm LF}/h^{-1}\Msun)} = 11.60^{+0.19}_{-0.20}$ and $11.50 \pm 0.16$ ($11.66 \pm 0.18$ and $11.62 \pm 0.19$), respectively. 
Moreover, all other deduced parameters, i.e., the satellite fraction, galaxy bias, effective halo mass, and fake fraction, show a similar level of agreement between the two models. 

Therefore, we can conclude that the presence of the Gaussian central occupation in less-massive haloes hardly affects the physical properties of haloes obtained through HOD-model fitting. 
We will use the results obtained from our HOD model without the central Gaussian occupation in the following section. 

\begin{table*}
\caption{The best-fitting and posterior mean HOD and its deduced parameters using our HOD model with and without the Gaussian central ELG occupation. }
\label{tab:hod_lf}
\begingroup
\renewcommand{\arraystretch}{1.18}
\begin{tabular}{llllllllllll} \hline \hline
& \multicolumn{5}{c}{Our HOD model} && \multicolumn{5}{c}{Our HOD model  (no Gaussian central occupation)} \\
 \cline{2-6}
 \cline{8-12}
& \multicolumn{2}{c}{NB816 $(z_{\rm eff} = 1.193)$} && \multicolumn{2}{c}{NB921 $(z_{\rm eff} = 1.471)$} && \multicolumn{2}{c}{NB816 $(z_{\rm eff} = 1.193)$} && \multicolumn{2}{c}{NB921 $(z_{\rm eff} = 1.471)$} \\ 
 \cline{2-3} \cline{5-6} \cline{8-9}\cline{11-12} 
\vspace{-0.2cm}
 & \multirow{2}{*}{Best fit} & Posterior && \multirow{2}{*}{Best fit} & Posterior && \multirow{2}{*}{Best fit} & Posterior && \multirow{2}{*}{Best fit} & Posterior \\ 
 & & mean && & mean & &  & mean & & & mean \\ 
 \hline
$\log_{10}{(L_{\rm Gauss})}^{a}$ & $42.06$ & $42.26^{+0.57}_{-0.62}$ && $41.48$ & $42.06^{+0.72}_{-0.67}$ && -- & -- && -- & -- \\
$\sigma_{\log{M_{h}}}$ & $0.29$ & $0.43^{+0.31}_{-0.31}$ && $0.24$ & $0.42^{+0.32}_{-0.31}$ && -- & -- && -- & -- \\
$\sigma_{\log{L}}$ & $0.09$ & $0.25^{+0.18}_{-0.18}$ && $0.52$ & $0.41^{+0.31}_{-0.30}$ && $0.18$ & $0.26^{+0.18}_{-0.19}$ && $0.05$ & $0.52^{+0.38}_{-0.37}$ \\
$\alpha_{\rm sat}$ & $0.65$ & $0.49^{+0.17}_{-0.19}$ && $0.69$ & $0.75^{+0.26}_{-0.30}$ && $0.67$ & $0.46^{+0.18}_{-0.20}$ && $0.78$ & $0.70^{+0.29}_{-0.33}$ \\
$\log_{10}{(L_{0})}^{a}$ & $41.49$ & $42.09^{+0.89}_{-0.85}$ && $41.19$ & $41.29^{+0.65}_{-0.68}$ && $42.03$ & $42.24^{+0.91}_{-0.92}$ && $42.11$ & $41.97^{+1.12}_{-1.01}$ \\
$\log_{10}{(M_{\rm trans})}^{b}$ & $11.41$ & $12.16^{+0.86}_{-0.86}$ && $11.24$ & $11.45^{+0.34}_{-0.36}$ && $12.08$ & $12.26^{+0.81}_{-0.86}$ && $11.38$ & $11.82^{+0.65}_{-0.61}$ \\
$\beta$ & $0.71$ & $0.77^{+0.17}_{-0.17}$ && $0.93$ & $0.69^{+0.25}_{-0.29}$ && $1.03$ & $0.88^{+0.24}_{-0.24}$ && $0.94$ & $1.32^{+0.44}_{-0.43}$ \\
$\gamma$ & $1.43$ & $1.57^{+0.68}_{-0.54}$ && $1.63$ & $2.12^{+0.50}_{-0.49}$ && $1.03$ & $1.67^{+0.74}_{-0.59}$ && $2.11$ & $2.19^{+0.49}_{-0.45}$ \\ 
$f_{\rm ELG}^{\rm min}$ & $0.70$ (fixed) & $0.70$ (fixed) && $0.70$ (fixed) & $0.70$ (fixed) && $0.70$ (fixed) & $0.70$ (fixed) && $0.70$ (fixed) & $0.70$ (fixed) \\
$f_{\rm ELG}^{\rm max}$ & $1.00$ (fixed) & $1.00$ (fixed) && $1.00$ (fixed) & $1.00$ (fixed) && $1.00$ (fixed) & $1.00$ (fixed) && $1.00$ (fixed) & $1.00$ (fixed) \\
$\alpha_{\rm ELG}$ & $2.47$ & $2.78^{+1.54}_{^1.58}$ && $3.33$ & $2.28^{+1.78}_{-1.68}$ && $4.46$ & $2.73^{+1.58}_{-1.61}$ && $2.66$ & $2.29^{+1.78}_{-1.70}$ \\
$\log_{10}{(M_{h, {\rm ELG}})}^{b}$ & $11.06$ & $11.37^{+0.27}_{-0.29}$ && $11.24$ & $11.86^{+0.63}_{-0.64}$ && $11.08$ & $11.42^{+0.30}_{-0.33}$ && $11.12$ & $11.89^{+0.62}_{-0.64}$ \\
$B_{\rm sat}$ & $0.76$ & $0.93^{+0.42}_{-0.44}$ && $0.39$ & $0.94^{+0.42}_{-0.43}$ && $0.39$ & $1.02^{+0.41}_{-0.43}$ && $0.86$ & $1.07^{+0.42}_{-0.42}$ \\
$\beta_{\rm sat}$ & $1.13$ & $1.31^{+0.48}_{-0.46}$ && $1.13$ & $1.14^{+0.61}_{-0.64}$ && $0.96$ & $1.35^{+0.47}_{-0.47}$ && $1.03$ & $1.11^{+0.62}_{-0.66}$ \\
$F_{\rm Gauss}$ & $0.26$ & $0.30^{+0.23}_{-0.23}$ && $0.62$ & $0.36^{+0.28}_{-0.27}$ && $0.00$ (fixed) & $0.00$ (fixed) && $0.00$ (fixed) & $0.00$ (fixed) \\
$F_{\rm erf}$ & $0.55$ & $0.71^{+0.21}_{-0.22}$ && $0.23$ & $0.71^{+0.21}_{-0.22}$ && $0.69$ & $0.75^{+0.19}_{-0.20}$ && $0.16$ & $0.71^{+0.22}_{-0.22}$ \\
$F_{\rm s}$ & $0.45$ & $0.59^{+0.28}_{-0.28}$ && $0.41$ & $0.54^{+0.31}_{-0.31}$ && $0.23$ & $0.56^{+0.29}_{-0.28}$ && $0.20$ & $0.53^{+0.31}_{-0.31}$ \\
$F_{\rm d}$ & $0.81$ & $0.55^{+0.32}_{-0.33}$ && $0.90$ & $0.52^{+0.34}_{-0.34}$ && $0.89$ & $0.59^{+0.30}_{-0.32}$ && $0.94$ & $0.54^{+0.33}_{-0.34}$ \\
$\log_{10}{(M_{\rm d})}$ & $12.11$ & $13.12^{+0.66}_{-0.70}$ && $13.04$ & $12.99^{+0.73}_{-0.75}$ && $12.43$ & $13.18^{+0.62}_{-0.69}$ && $12.23$ & $13.09^{+0.66}_{-0.67}$ \\
$\sigma_{\rm d}$ & $0.19$ & $0.53^{+0.32}_{-0.32}$ && $0.28$ & $0.52^{+0.32}_{-0.32}$ && $0.22$ & $0.52^{+0.32}_{-0.32}$ && $0.33$ & $0.52^{+0.32}_{-0.32}$ \\ \hline  \vspace{-0.1cm}
\multirow{2}{*}{$\chi^{2}/{\rm dof}$} & $6.93/15$ & $12.53/15$ && $12.50/14$ & $17.88/14$ && $6.59/18$ & $23.52/18$ && $12.69/17$ & $18.68/17$ \\
& $= 0.46$ & $= 0.84$ && $ = 0.89$ & $ = 1.28$ && $ = 0.37$ & $ = 1.31$ && $ = 0.75$ & $ = 1.10$ \\
$\chi^{2}_{\rm ACF}$ & $3.85$ & $4.31$ && $7.24$ & $7.06$ && $3.33$ & $14.99$ && $7.03$ & $7.12$ \\
$\chi^{2}_{\rm LF}$ & $3.08$ & $8.23$ && $5.27$ & $10.82$ && $3.25$ & $8.53$ && $5.67$ & $11.57$ \\
$\log_{10}{(\bar{n_{\rm ELG}})}^{c}$ & $-1.88$ & $-1.95^{+0.13}_{-0.13}$ && $-1.92$ & $-2.20^{+0.15}_{-0.15}$ && $-1.96$ & $-2.03^{+0.13}_{-0.13}$ && $-2.17$ & $-2.32^{+0.11}_{-0.11}$ \\
$f_{\rm fake}$ &$0.02$ & $0.08^{+0.05}_{-0.05}$ && $0.06$ & $0.14^{+0.07}_{-0.07}$ && $0.01$ & $0.08^{+0.05}_{-0.05}$ && $0.06$ & $0.15^{+0.07}_{-0.07}$ \\
$\log_{10}{(\Meff)}^{b}$ & $12.31$ & $12.30^{+0.08}_{-0.08}$ && $12.22$ & $12.33^{+0.11}_{-0.11}$ && $12.28$ & $12.30^{+0.08}_{-0.08}$ && $12.26$ & $12.33^{+0.11}_{-0.11}$ \\
$\log_{10}{(M_{\rm min}^{\rm LF})}^{b}$ & $11.37$ & $11.60^{+0.19}_{-0.20}$ && $11.32$ & $11.66^{+0.18}_{-0.19}$ && $11.47$ & $11.50^{+0.16}_{-0.16}$ && $11.04$ & $11.62^{+0.19}_{-0.19}$ \\
$\bg$ & $1.41$ & $1.44^{+0.05}_{-0.05}$ && $1.57$ & $1.71^{+0.09}_{-0.09}$ && $1.42$ & $1.45^{+0.05}_{-0.05}$ && $1.56$ & $1.72^{+0.09}_{-0.09}$ \\
$\fsat$ & $0.36$ & $0.31^{+0.08}_{-0.08}$ && $0.38$ & $0.20^{+0.06}_{-0.06}$ && $0.31$ & $0.30^{+0.08}_{-0.08}$ && $0.38$ & $0.18^{+0.06}_{-0.06}$ \\\hline
\multicolumn{12}{l}{\footnotesize$^a$ Parameters regarding luminosities are in unit of erg/s in logarithmic scale. } \\
\multicolumn{12}{l}{\footnotesize$^b$ Parameters regarding halo masses are in unit of $h^{-1}\Msun$ in logarithmic scale. } \\
\multicolumn{12}{l}{\footnotesize$^c$ Number densities of ELGs are in unit of $(h^{-1}{\rm Mpc})^{-3}$ in logarithmic scale. } \\
\end{tabular}
\endgroup
\end{table*}

\section{Discussions} \label{sec:discussion}
\subsection{Connecting $\oii$ emitting galaxies to other populations} \label{subsec:halo_mass_evolution}
\subsubsection{Connecting $\oii$ emitters to other populations via extended Press--Schechter halo mass growth} \label{subsubsec:halo_mass_evolution}
We trace the redshift evolution of characteristic dark halo masses of our $\oii$ NB816 and NB921 samples and compare them to those of various galaxy populations calculated by previous HOD studies. 
We adopt $M_{\rm min}^{\rm LF}$ parameter as the characteristic dark halo mass of our $\oii$ emitters and $\Mmin$ parameter of other galaxy populations analysed using the classical HOD model proposed by \citet{zheng05} to match the definition of the dark halo mass. 

\begin{figure}
\includegraphics[width=\linewidth]{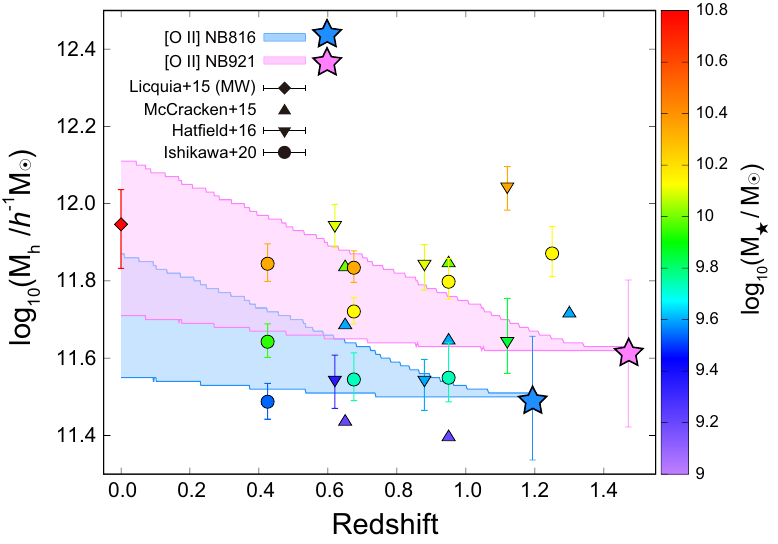}
\caption{Halo masses and their redshift evolution of our $\oii$-emitting galaxies observed through NB816 (blue) and NB921 (red) filters. Shaded regions represent the $1\sigma$ confidence intervals of halo mass evolution predicted by the extended Press--Schechter formalism. For comparison, halo masses of photo-$z$ selected galaxies at various redshifts calculated by HOD-model analysis \citep{mccracken15,hatfield16,ishikawa20} and the virial mass of the Milky Way \citep{licquia15} are also plotted. Colours other than those for our $\oii$ emitter samples observed through NB816 and NB921 filters indicate the stellar mass threshold of each galaxy sample. Note that the $\Mmin$ parameter of the HOD model is adopted as the halo mass in this figure. }
\label{fig:mh_eps}
\end{figure}

Figure~\ref{fig:mh_eps} illustrates the redshift evolution of dark halo masses evaluated in this study. 
It is predicted using an extended Press--Schechter formalism \citep[EPS;][]{bond91,bower91,lacey93}\footnote{https://sci.nao.ac.jp/MEMBER/hamana/OPENPRO/index.html}, which is an extension of the classical excursion set approach \citep{press74}.
The EPS formalism takes into account the probability distribution function (PDF) of halo mass, and the PDF is proportional to the conditional probability that a given dark halo at a certain redshift will be incorporated into a more massive dark halo at a lower redshift. 
We estimate the $1\sigma$ confidence intervals of the halo mass PDFs at each redshift \citep[see][for more details]{hamana06}. 

Halo masses of various galaxy populations are also plotted in Figure~\ref{fig:mh_eps} to investigate the potential descendants of our $\oii$-emitting galaxies from the perspective of halo mass evolution. 
The plotted halo masses are derived from HOD analyses of galaxy samples selected based on photometric redshifts and stellar-mass thresholds. 
It is important to note that the halo masses shown in Figure~\ref{fig:mh_eps} \citep{mccracken15,hatfield16,ishikawa20} are not confined to specific galaxy colours or types, i.e., they include both blue and red galaxies. 
These studies fitted their ACFs using the HOD model by \citet{zheng05}, incorporating number density constraints, and the $\Mmin$ parameter of the HOD model is adopted as the halo mass. 
It should be noted that the $1\sigma$ confidence intervals of halo masses from HOD studies other than our results are quite small since the $\Mmin$ parameter is strongly constrained by the number densities of galaxies. 
However, the $1\sigma$ confidence levels of both of our $\oii$ emitter samples are relatively large compared to other studies because the halo masses are plotted using the relation given in equation~\ref{eq:Mmin_elg} that is introduced to match the definition of $\Mmin$ in the HOD model of \citet{zheng05}. 
We also include the virial mass of the Milky Way, estimated from the distributions of globular cluster systems observed by {\it Gaia} and the {\it Hubble} Space Telescope, assuming the concentration--virial mass relation \citep{posti19}. 
For the Milky Way's stellar mass, we adopt the estimate by \citet{licquia15}, which was derived using a hierarchical Bayesian statistical approach.

According to the EPS approach, halo masses of our $\oii$-emitting galaxy samples reach $\sim 10^{12} \Msun$ in the present-day Universe, which corresponds to the mass range of the Milky Way galaxy. 
Notably, the $\oii$ emitters observed at $z \sim 1.4$ grow their halo masses through merging, and the $1\sigma$ confidence interval of the halo mass at $z = 0.0$ fully encompasses the estimated virial mass of the Milky Way. 
In addition to the halo mass evolution, the evolutionary connection of stellar masses across redshifts also provides important insights into the relationship between galaxy populations. 
The minimum (corresponding to the stellar mass threshold) and median stellar masses of our $\oii$ NB921 sample are $8.07$ and $10.13$ $\Msun$ in logarithmic scale, respectively. 
Therefore, the stellar mass evolution along the EPS-predicted evolutionary path of the halo mass is consistent from the $\oii$ NB921 sample to the Milky Way. 
This result strongly suggests that our $\oii$ NB921samples can be descendants or building blocks of Milky-Way-like galaxies in present-day Universe. 

The $\oii$-emitting galaxies observed at $z \sim 1.2$, on the other hand, are expected to evolve into galaxies whose total masses are slightly lower than that of the Milky Way at $z \sim 0.0$, although some highly evolved ones may reach the mass range of the Milky Way. 
The minimum and median stellar masses of the $\oii$ NB816 sample are evaluated as $8.05$ and $9.92$ in logarithmic scale, respectively. 
Therefore, the stellar mass evolution of galaxy populations along the EPS-predicted halo mass evolutionary track appears to be consistent from $z \sim 1.2$ to $z = 0.0$; i.e., the stellar masses of $\oii$ NB816 galaxies rapidly grow through bursty star formation activity and then continuously increase their stellar masses through star formation and/or merging, reaching $\sim 10^{10}\Msun$ at $z \sim 0.4$. 
According to the halo and stellar mass evolution, most of the $\oii$-emitting galaxies observed with the NB816 filter are expected to evolve into galaxies slightly smaller than the Milky Way at $z = 0.0$. 

\subsubsection{Evolution of satellite ELGs over $z<2$} \label{subsubsection:satellite_fraction}
\begin{figure}
\includegraphics[width=\linewidth]{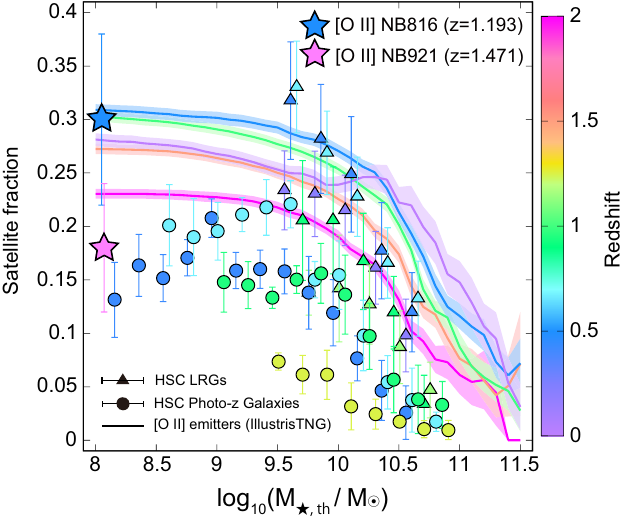}
\caption{Comparison of the satellite fractions for our $\oii$-emitting galaxies at $z=1.193$ (blue star) and $z=1.471$ (red star) with other galaxy populations at various redshifts as a function of stellar mass limit. The solid lines with shaded regions show the theoretical predictions for $\oii$ emitters from the IllustrisTNG simulation at $z = 0.0$ (purple), $0.5$ (blue), $1.0$ (green), $1.5$ (orange), and $2.0$ (magenta) \citep{springel18,nelson19,osato23}. Observational results are also shown for two galaxy populations selected in the HSC SSP Wide layers: HSC CAMIRA LRGs at $0.1 \leq z \leq 1.25$ \citep[upward triangles;][]{ishikawa21,ishikawa24} and HSC photo-$z$ galaxies at $0.3 \leq z \leq 1.4$ \citep[circles;][]{ishikawa20}. The colours of the TNG $\oii$-emitters, HSC CAMIRA LRGs, and HSC photo-$z$ galaxies indicate their respective redshifts (see the colour bar on the right). The observational errors correspond to the $1\sigma$ uncertainties in the posterior mean HOD parameters, whilst those for IllustrisTNG are estimated via jackknife resampling of the simulated ELG catalogues. The data points for the HSC CAMIRA LRG and HSC photo-$z$ galaxy samples are slightly shifted along the $x$-axis for visual clarity. }
\label{fig:comp_fsat}
\end{figure}

In this section, we shift our focus from the central $\oii$ emitters discussed in Section~\ref{subsubsec:halo_mass_evolution} to the satellite $\oii$ emitters, exploring their redshift evolution and their potential evolutionary relationships with other galaxy populations, in particular LRGs. 
Figure~\ref{fig:comp_fsat} compares our measurements of the satellite fraction $\fsat$ for $\oii$ emitters at $z=1.193$ and $z=1.471$ (blue and red stars) to several relevant populations as a function of stellar mass limit. 
Satellite fraction of our ELGs are compared to those of the theoretical predictions for $\oii$ emitters from the IllustrisTNG simulation \citep[solid lines][; hereafter TNG $\oii$ emitters]{springel18,nelson19,osato23}, as well as the observational results for two galaxy populations selected in the HSC SSP Wide layers: HSC CAMIRA LRGs at $0.1 \leq z \leq 1.25$ \citep[upward triangles;][]{ishikawa21,ishikawa24} and the HSC photo-$z$ galaxies at $0.3 \leq z \leq 1.4$ \citep[circles;][]{ishikawa20}. 
The shaded regions of IllustrisTNG predictions are evaluated via the JK resampling of the simulated ELG catalogues ($125$ subsamples for each ELG catalogue from the IllustrisTNG), whilst the errors of the observational results correspond to the $1\sigma$ uncertainties in the posterior mean HOD parameters. 
In constructing the TNG $\oii$ emitter catalogues, we convert the line luminosities of $\oii \lambda$3727 forbidden-line doublet into line fluxes at each redshift, and then pose the flux limit $f_{\oii} \geq 3.0 \times 10^{-17}$ [erg/s/cm$^{2}$] similar to our HSC $\oii$ catalogues. 

Our $\oii$ emitter at $z=1.193$ is broadly consistent with the TNG $\oii$ emitters prediction, indicating that the simulation can capture the fraction of star-forming satellites reasonably well at this epoch. 
However, at $z=1.471$, we find that our measured satellite fraction is more than $1\sigma$ lower than the prediction of the IllustrisTNG at $z=1.5$ (shown by the orange solid line in the plot). 
This discrepancy can be attributed to the IllustrisTNG simulation may overestimate the abundance of satellite galaxies that retain active star formation at such high redshifts, perhaps because its feedback mechanisms are insufficiently strong or because its dust attenuation prescriptions do not fully obscure satellites with ongoing star formation. 
Under these conditions, the IllustrisTNG would predict a higher number of luminous satellite ELGs than is observed. 
Alternatively, observational biases, such as missing dusty satellites in our narrow-band selection, might contribute to reducing the observed satellite fraction for our ELG samples, though further comparisons with deeper surveys are needed to quantify this effect. 

Interestingly, IllustrisTNG also shows that the satellite fraction among $\oii$ emitters peaks at $z=0.5-1.0$ and declines towards $z=0$. 
This trend suggests that satellite ELGs originally accreted with sufficient gas to sustain star formation can remain ELGs for a certain period of cosmic time, but ultimately cease forming stars through mechanisms such as gas depletion or starvation \citep{larson80,wright22}, ram-pressure stripping \citep{abadi99}, galaxy harassment \citep{moore96}, and tidal interactions \citep{bekki99} by the present day. 
Consequently, such an environmental quenching effect \citep[e.g.,][]{peng12,gonzalez22} efficiently suppresses star formation in satellite galaxies, causing the number of ELG satellites to drop at $z<0.5$. 

In addition to environmental processes, correlations between central and satellite galaxies can further drive down the ELG satellite fraction towards $z=0$. 
Observational and theoretical studies have shown that the star formation activity of satellites often mirrors that of their centrals, a phenomenon known as $1$-halo galactic conformity \citep[e.g.][]{kawinwanichakij16,berti17}. 
During the interval $z \sim 0.5$ to $z=0$, both observations and simulations indicate that the quenched fraction of central galaxies rises sharply, for instance due to cumulative feedback processes and the exhaustion of available cold gas \citep[e.g.][]{donnari21,arango24}. 
As centrals transition into a quenched state, the same physical conditions, such as reduced gas supply \citep{mcgee14}, can accelerate quenching among their satellites, thus reducing the population of ELG satellites by $z=0$. 
Acting in tandem with direct environmental processes (e.g. tidal stripping, ram-pressure effects), these factors may collectively account for the pronounced decline in the ELG satellite fraction at late cosmic times. 

Turning first to a comparison with the HSC photo-$z$ sample, we find that the satellite fraction among $\oii$ emitters at $z \approx 1.2-1.5$ is somewhat higher ($\fsat \approx 0.2-0.4$) than the satellite fractions typically measured for photo-$z$ galaxies at similar or lower redshifts ($\fsat<0.15$, \citealt{ishikawa20}). 
This difference likely arises because ELG selection, particularly via $\oii$, preferentially identifies actively star-forming satellite galaxies residing in relatively low-mass dark matter haloes, which, due to their strong emission lines, would otherwise be less conspicuous in a purely luminosity-selected or quiescence-dominated sample. 
Such satellite galaxies typically maintain high specific star-formation rates for a limited cosmic interval. 
Over time, environmental processes with central galaxies and satellite--satellite mergers \citep{simha09,bahe19} can reduce their star formation activity and visibility as ELGs. 
Consequently, the relatively large satellite fractions observed for ELGs at $z \approx 1.2-1.5$ naturally decline toward the lower satellite fractions observed in photo-$z$ galaxy samples at later cosmic epochs. 

In contrast to the photo-$z$ results, the HSC LRG sample exhibits a higher satellite fraction of approximately 0.3 at $z<0.5$, comparable to the fractions measured for $\oii$ emitters at $z \approx 1.2-1.5$. 
Past HOD analyses focusing only on passive galaxy populations \citep[e.g.][]{coupon12,mccracken15} also find similarly elevated satellite fractions. 
This similarity may suggest an evolutionary link between these populations, in which high-$\fsat$ star-forming ELG satellites at higher redshifts evolve through environmental quenching processes \citep[e.g.,][]{delucia07,vdbosch08} into passive or red-sequence galaxies such as LRGs at lower redshifts. 
Such an evolutionary connection also indicates that massive cluster-scale systems observed at low redshifts, populated predominantly by LRG-type galaxies \citep[e.g.,][]{rykoff14,oguri18}, likely originated as structures hosting numerous ELG satellites at earlier cosmic epochs. 

In this evolutionary framework, environmental processes such as tidal stripping, ram-pressure quenching, and minor mergers are all expected to transform a population of actively star-forming satellites into fewer, more massive, or quenched descendants by $z \sim 0$. 
IllustrisTNG captures some of these processes by exhibiting a peak in the ELG satellite fraction at intermediate redshifts before declining by the local Universe, although the precise agreement with observations remains imperfect, especially at  $z>1.4$. 
Future improvements in both feedback physics and dust modelling in the hydrodynamical simulations, and more comprehensive narrow-band and spectroscopic observations of high-redshift star-forming satellites, are crucial for resolving these residual discrepancies. 
Ultimately, linking the star-forming satellite fractions at high redshifts to their low-redshift descendants will shed further light on how galaxy evolution in dense environments proceeds from $z \sim 2$ to the present-day Universe. 

\subsection{Co-evolutions of $\oii$-emitting galaxies between baryonic properties and host dark haloes}
\subsubsection{Luminosity-to-halo mass relation of $\oii$-emitting galaxies} \label{subsubsec:lhmr}

Our HOD model incorporates the relationship between line luminosity and dark halo mass of central ELGs ($f_{\rm LHMR}(\Mh)$, see equation~\ref{eq:lhmr}). 
In this section, we check the LHMR of our $\oii$-emitting galaxies evaluated through the HOD-model fitting and then compare them with the observed results in literature. 

\begin{figure}
\includegraphics[width=\linewidth]{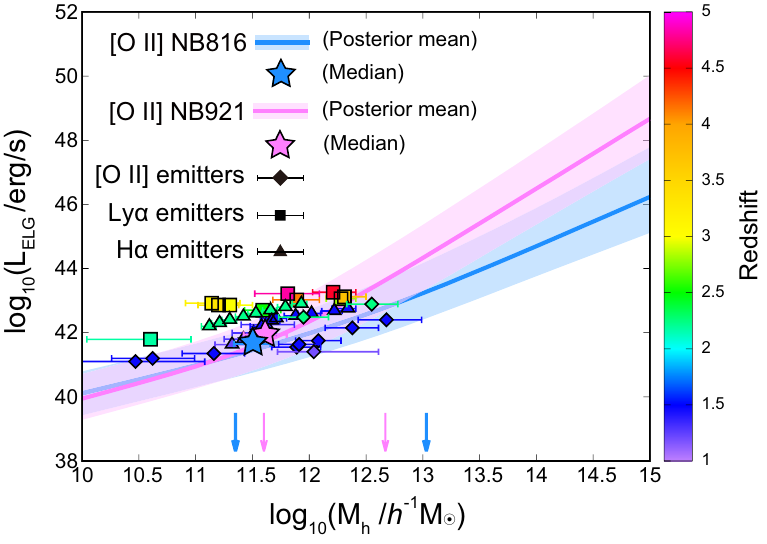}
\caption{This figure illustrates the relationship between the line luminosities of ELGs and their host halo masses, referred to as the LHMR in this study. Our $\oii$ emitter samples observed through NB816 (blue, $z=1.193$) and NB921 (red, $z=1.471$) are presented with both posterior mean results (solid lines with shaded regions) and median results (star symbols). The posterior mean LHMRs are derived from equation~\ref{eq:lhmr} using corresponding HOD parameters, whilst the median LHMRs are calculated using the median line luminosities of our observed samples and $M_{\rm min}^{\rm LF}$ parameters. The shaded regions represent the $16$th and $84$th percentiles of the LHMRs, computed by varying the HOD parameters within $1\sigma$ confidence levels. We also show the LHMRs of ELGs from the literature: $\oii$ emitters \citep[diamonds;][]{khostovan18,okumura21}, LAEs \citep[squares;][]{kusakabe18,khostovan19}, and HAEs \citep[triangles;][]{cochrane17,clontz22}, with their redshifts indicated by the colour bar. The downward arrows above the horizontal axis represent the expected minimum/maximum halo mass derived from the observed minimum/maximum $\oii$ line luminosities of our samples using the posterior mean LHMRs.}
\label{fig:lhmr}
\end{figure}

Figure~\ref{fig:lhmr} shows the line luminosities of central ELGs as a function of halo mass. 
For our $\oii$ emitters, we present both the posterior mean results of $f_{\rm LHMR}(\Mh)$ (solid lines with shaded regions) and the median results, calculated based on the median observed $\oii$ luminosities of our samples and $M_{\rm min}^{\rm LF}$ parameters (star symbols with error bars). 
We also include the $\Mh$--$L_{\rm line}$ relations of ELGs from various studies: $\oii$ emitters \citep[][diamonds]{khostovan18,okumura21}, Lyman $\alpha$ emitters \citep[LAEs;][squares]{kusakabe18,khostovan19}, and H$\alpha$ emitters \citep[HAEs;][triangles]{cochrane17,clontz22}. 
The observational results from previous studies utilised for comparison in this section are obtained from narrow-band imaging surveys, covering a wide range of flux and magnitude thresholds as well as redshift intervals that broadly overlap with those of our samples, ensuring a meaningful and consistent context for the comparisons presented below. 

Our results align well with other studies at similar redshifts $(1 \lesssim z \lesssim 2)$, showing that both the posterior mean and median of LHMRs for our $\oii$-emitting galaxies are consistent. 
This supports the validity of the formulated LHMR as a double power law as a function of halo mass (equation~\ref{eq:lhmr}), at least within the redshift ranges of our $\oii$ samples and their halo mass ranges, which are indicated by downward arrows in Figure~\ref{fig:lhmr}.
However, the bright ELGs corresponding to massive haloes are extremely rare, making it statistically challenging to constrain the bright end of the LHMRs in both our study and the literature. 
Thus, further observational confirmation of the LHMR at the massive end, particularly from upcoming extensive spectroscopic surveys targeting ELGs, is crucial. 

We observe a slight evolution of the LHMR with redshift; specifically, the line luminosities of ELGs at fixed halo masses increase with redshift and this redshift evolution is clear for $\Mh \lesssim 10^{12} h^{-1}\Msun$ haloes. 
This trend may be due to efficient star formation at high redshift, driven by higher baryon accretion rates \citep[e.g.,][]{rasera06,fakhouri10} and a greater cold gas fraction \citep[e.g.,][]{vdvoort11} compared to lower redshifts. 
Additionally, feedback mechanisms such as supernovae and AGN might be less effective at higher redshifts, allowing for more uninterrupted star formation \citep[e.g.,][]{roos15}. 
Consequently, young galaxies at high-$z$ can burstingly form stars against a backdrop of abundant cold gas, resulting in stronger line luminosities than ELGs at low-$z$ with similar halo masses. 
Conversely, at fixed redshift, we find no evidence of different LHMRs among ELGs with different metal lines.

The transition halo masses of the LHMRs $\Mtrans$ of our $\oii$ samples are $\log_{10}(\Mtrans/h^{-1}\Msun) = 12.26 ^{+0.81}_{-0.86}$ and $11.82 ^{+0.65}_{-0.61}$ for NB816 and NB921 $\oii$ emitters, respectively. 
Although the errors of the transition halo masses are relatively large due to significant uncertainties in our observed LFs, the relationship between ELG line luminosities and host halo mass clearly changes around $\Mtrans$. 
This change can be attributed to differing formation scenarios of ELGs, which vary depending on the transition halo mass. 
Previous studies have reported that galaxies with low stellar masses at $z<2$ tend to exhibit high specific star-formation rates (sSFR) with low metallicity, indicating ongoing active star formation in young galaxies \citep[e.g.,][]{ilbert15,lin23}. 
In particular, low-mass ELGs at $z<1.5$ show bursty star-formation histories \citep[e.g.,][]{guo16,atek22}. 
Therefore, central ELGs residing in less-massive haloes ($\Mh \lesssim \Mtrans$) are thought to be young galaxies in a bursty star-forming phase. 

On the other hand, star formation in galaxies within massive haloes ($\Mh \gtrsim \Mtrans$) is likely to have passed its peak \citep[e.g.,][]{leauthaud12,behroozi13,behroozi19,ishikawa20}. 
The main origins of metal emission lines from massive ELGs may differ from the star-formation activity observed in low-mass ELGs. 
The $\oii$ emission lines can be excited by AGN activities, which are more likely to occur in massive haloes \citep{vitale13,maddox18,vietri22}. 
Moreover, massive haloes tend to reside in high-density regions such as knots in the cosmic web, where galaxy interactions and mergers are more frequent. 
Tidal forces induced by these interactions and mergers can perturb the dynamics of the interstellar medium and compress it. 
This process triggers shock excitation, which induces strong ionised line emissions \citep[e.g.,][]{roche16,kewley19} and can stimulate AGN activities \citep[e.g.,][]{gao20,steffen23}. 
Considering these results, massive $\oii$ emitters are likely to be relatively old, metal-rich galaxies located in high-density regions. 
A possible interpretation is that their $\oii$ emissions are thought to be largely attributable to shock excitations and AGN activities rather than recent bursty star formation, although future spectroscopic observations are required to directly confirm these mechanisms. 

\subsubsection{Baryon conversion efficiencies of $\oii$-emitting galaxies} \label{subsubsec:bce}

Baryon conversion efficiency (BCE) describes the fraction of formed star to accreted baryon onto galaxies as a function of halo mass. 
The BCE can be expressed as follows:
\begin{equation}
{\rm BCE}(\Mh, z) = \left. \frac{d\Mstar}{dt} \middle/ \frac{dM_{\rm b}}{dt} \right., 
\label{eq:bce}
\end{equation}
where $\frac{d\Mstar}{dt}$ and $\frac{dM_{\rm b}}{dt}$ represent the SFR and baryon accretion rate (BAR), respectively. 
The BAR can be evaluated as follows:
\begin{equation}
\frac{dM_{\rm b}}{dt} = f_{\rm b} \times \frac{d\Mh}{dt}, 
\label{eq:bar}
\end{equation}
where $f_{\rm b}$ denotes the baryon fraction and $\frac{d\Mh}{dt}$ is the mass accretion rate of dark haloes. 
In our cosmological model, the baryon fraction can be estimated as $f_{\rm b} = \Omega_{\rm b} / \Omega_{\rm m} \sim 0.156$. 
The mass accretion rate of dark halo is modelled by \citet{fakhouri10}, who evaluated the mean and median mass growth rates of dark haloes using the dataset of Millennium and Millennium-II simulations \citep{springel05,boylan09}, and we use the median mass growth rate to predict the redshift evolution of dark halo masses of our $\oii$-emitting galaxies. 

We calculate the BCEs of our $\oii$-emitting galaxies by two ways: averaged BCEs evaluated by the posterior mean LHMR derived from the HOD-model analyses and median BCEs computed using the median values of observed SFRs of our samples. 
In evaluating the posterior mean BCEs, the SFRs of our $\oii$-emitting galaxies are converted from the $\oii$ line luminosities $L_{\oii}$ using the relation presented by \citet{sobral12} as follows:
\begin{equation}
\frac{d\Mstar}{dt} \left(\Msun/{\rm yr}\right) = 1.4 \times 10^{-41} L_{\oii} \left({\rm erg/s}\right),
\label{eq:sfr-l}
\end{equation}
which builds on the standard calibration by \citet{kennicutt98} and \citet{madau98}. 
This conversion from line luminosities to SFRs is widely used in many observational studies and has been tested across multiple wavelengths (e.g.\ UV, IR, and other emission lines). 
On the other hand, the median BCEs are estimated using the SFRs of each ELG that are evaluated by the SED-fitting procedures using {\sc Mizuki} code \citep{tanaka15}. 

\begin{figure}
\includegraphics[width=\linewidth]{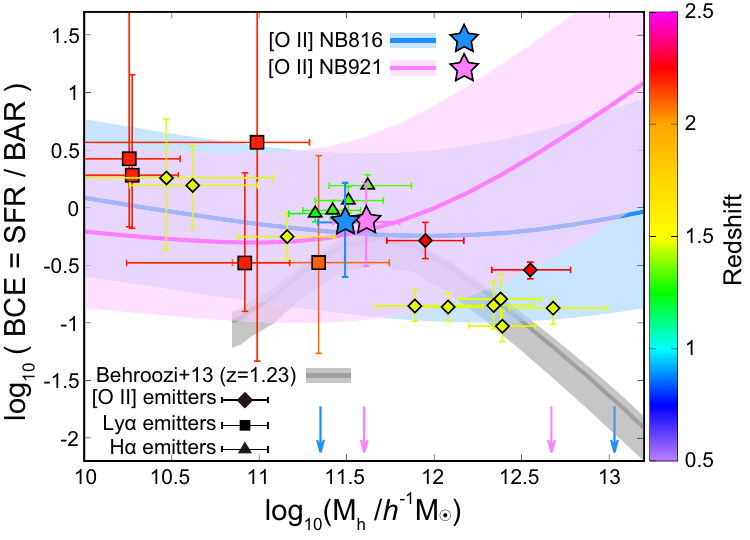}
\caption{The baryon conversion efficiencies of ELGs as a function of halo mass. The results for our $\oii$ samples are plotted in the same manner as in Figure~\ref{fig:lhmr}. We also display the theoretical predictions presented by \citet{behroozi13} at $z=1.23$, as well as other observational results of the $\oii$ emitters \citep[diamonds;][]{khostovan18}, the LAEs \citep[squares;][]{guaita11,kusakabe18}, and the HAEs \citep[triangles;][]{clontz22}, with their redshifts indicated by the colour bar. }
\label{fig:bce}
\end{figure}

Figure~\ref{fig:bce} illustrates the BCEs of central ELGs as a function of halo mass. 
In this plot, we also include the theoretical prediction using the abundance matching technique from \citet{behroozi13} at $z=1.23$, as well as other observational results for the $\oii$ emitters \citep[diamonds;][]{khostovan18}, the LAEs \citep[squares;][]{guaita11,kusakabe18}, and the HAEs \citep[triangles;][]{clontz22}. 
The BCEs from \citet{khostovan18} and \citet{clontz22} are calculated using the SFRs converted from the line luminosities via the empirical relation in equation~\ref{eq:sfr-l}. 

Focusing on the BCEs of massive haloes $(\Mh \gtrsim 10^{11.6}h^{-1}\Msun)$, our posterior mean results show an opposite trend compared to the theoretical prediction by \citet{behroozi13}. 
Specifically, our posterior mean BCEs have minimum values at $\Mh \sim 10^{12}h^{-1}\Msun$ and increase towards both less-massive and massive ends. 
This trend originates from the SFRs of our posterior mean results, which are derived from the LHMRs assumed to be a convex downward function. 
As mentioned in Section~\ref{subsubsec:lhmr}, it is unclear whether the LHMR maintains the same power-law index at the massive end. 
Currently available observational data, including our HSC SSP Deep/UltraDeep layers, are quite small in bright ELGs, making it difficult to statistically constrain it. 
Therefore, discussing the discrepancies between our results and \citet{behroozi13} at the massive end is beyond the scope of this paper. 

Subsequently, the halo masses of the median BCEs of our $\oii$-emitting galaxy samples nearly correspond to the predicted peak halo mass of BCEs presented by \citet{behroozi13}. 
At this halo mass scale $(\Mh \sim 10^{11.6}h^{-1}\Msun)$, our median BCEs and the results of HAEs \citep{clontz22} slightly exceed the theoretical predictions, albeit within the $1\sigma$ confidence intervals for our results. 
The BCEs from \citet{behroozi13} are calculated using observational constraints, particularly the cosmic SFR (cSFR) density and the sSFR, obtained from various types of galaxy samples. 
At $z>1$, large fractions of these observational constraints are traced by UV, IR, and radio fluxes, with H$\alpha$ emission lines contributing relatively little to constraining the BCEs. 
The smaller amplitudes of the BCEs from the theoretical prediction can be attributed to the difference in star formation tracers, implying that the star-forming activities of the ELGs are quite large compared to other star-forming galaxies. 

Finally, focusing on the less-massive scales $(\Mh \lesssim 10^{11.6}h^{-1}\Msun)$, our posterior mean results apparently exceed the theoretical prediction of \citet{behroozi13}. 
Interestingly, the trend of halo mass dependence of BCEs calculated by other observational studies shows excellent consistency with our results. 
ELGs hosted by less-massive haloes, especially at $\Mh \lesssim 10^{11.0}h^{-1}\Msun$, gradually increase with decreasing halo mass, potentially reflecting efficient star formation fuelled by cold-mode gas accretion from the circumgalactic medium (CGM). 
Indeed, previous studies have indicated that the CGM in these haloes predominantly consists of cold gas ($T<10^{5}$ K), which can directly feed star formation via filamentary structures \citep[e.g.,][]{ford14,tumlinson17,keres05,vdvoort12}. 
In addition, the era during which ${\rm BCE}>10^{0}$ is observed ($1.5 \lesssim z \lesssim 2.2$) coincides with the epoch when the cosmic star formation rate (cSFR) density peaks \citep{madau98,madau14}. 
During this epoch, low-mass ELGs may experience brief but intense starburst episodes, temporarily elevating their BCE values significantly above the cosmic average. 
However, strong supernova-driven galactic winds \citep[e.g.,][]{shankar06,muratov15} eventually expel large fractions of their gas reservoirs from the CGM, leading to suppression of star formation. 
This suppression naturally reduces the BCEs at subsequent epochs, guiding the evolution of ELGs towards our posterior mean results (${\rm BCE} \sim 10^{-0.2 - 0.0}$) at later cosmic times ($z=1.1-1.4$). 
Future observational campaigns investigating CGM properties and gas outflows around these galaxies will help clarify and validate this evolutionary scenario. 

\subsubsection{Stellar-to-halo-mass relationship of $\oii$-emitting galaxies} \label{subsubsec:shmr}

The SHMR is defined as the fraction between stellar masses and halo masses, which corresponds to the star-formation efficiency of galaxies, as a function of halo mass. 
Previous theoretical and observational studies have revealed that the most efficient halo mass of forming stars is almost constant at $\Mh \sim 10^{12} h^{-1}\Msun$ up to $z \sim 5$, but it slightly increases with redshift that is consistent with the downsizing of galaxies \citep[e.g.,][]{leauthaud12,behroozi13,ishikawa17,ishikawa20,legrand19}. 

We calculate the posterior mean and median SHMRs of our $\oii$-emitting samples, similar to the case of the BCEs. 
In calculating the SHMRs of the posterior mean results, the stellar masses of $\oii$-emitting galaxies are evaluated from the SFRs, which are also converted from the $\oii$ line luminosities (see Section~\ref{subsubsec:bce}), assuming the main sequence of star-forming galaxies. 
We adopt the best-fitting results of the main sequence relation presented by \citet{speagle14} as follows:
\begin{equation}
\log_{10}(\frac{d\Mstar}{dt}) = (0.84 - 0.026 t)\log_{10}(\Mstar) - (6.51 - 0.11 t),
\label{eq:ms}
\end{equation}
where $t$ denotes cosmic time in unit of Gyr, whilst the unit of the SFR in the above equation is still $\Msun$/yr. 
In our Planck 2018 cosmological model, cosmic age at $z=1.193$ ($\oii$ NB816) and $z=1.471$ ($\oii$ NB921) can be estimated as $t = 5.140$ and $4.337$ Gyr, respectively. 
It is noted that the scatter in the fitted main sequence relation is not considered in our analysis since the SFRs evaluated from the line luminosities already omit the effects of both scatter and photometric errors. 
The median SHMRs, on the other hand, are defined by the ratios between the median values of observed stellar masses estimated by the SED-fitting technique and $M_{\rm min}^{\rm LF}$. 

\begin{figure}
\includegraphics[width=\linewidth]{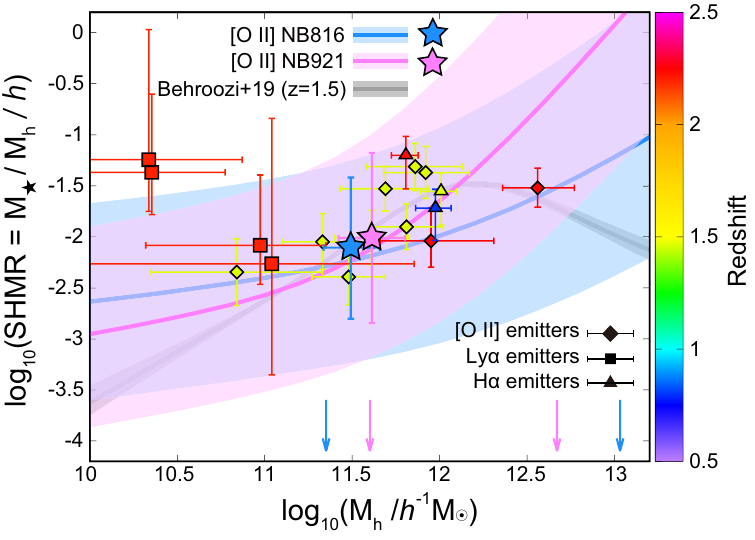}
\caption{The stellar-to-halo mass relationship of ELGs as a function of halo mass. The results for our $\oii$ samples are plotted in the same manner as in Figure~\ref{fig:lhmr}. We also show the SHMRs of the theoretical prediction of central star-forming galaxies at $z\sim1.5$ presented by \citet{behroozi19}, as well as those of other observational results of the $\oii$ emitters \citep[diamonds;][]{khostovan18}, the LAEs \citep[squares;][]{kusakabe18}, and the HAEs \citep[triangles;][]{cochrane18}, with their redshifts indicated by the colour bar. }
\label{fig:shmr}
\end{figure}

Figure~\ref{fig:shmr} displays the SHMRs of central ELGs as a function of halo mass. 
We also include the theoretical prediction from \citet{behroozi19} at $z=1.5$, derived using observational constraints of central star-forming galaxies at this epoch, as well as other observational results for the $\oii$ emitters \citep[diamonds;][]{khostovan18}, the LAEs \citep[squares;][]{kusakabe18}, and the HAEs \citep[triangles;][]{cochrane18}. 
For the same reason as in Section~\ref{subsubsec:bce} and Figure~\ref{fig:bce}, we only discuss the SHMRs at $\Mh \lesssim 10^{11.6} h^{-1}\Msun$. 

At $\Mh \sim 10^{11.6} h^{-1}\Msun$, the SHMRs from both our posterior mean and median results show good agreement with the theoretical prediction of \citet{behroozi19}. 
Since the BCEs derived from our results slightly exceed the theoretical model, our $\oii$ emitter samples are likely nearing the end of their rapid stellar mass assembly phase. 
Such efficient stellar mass assembly in galaxies has been suggested to depend strongly on galaxy morphology and disc gravitational instabilities, which regulate stellar mass fractions across diverse galaxy populations \citep[e.g.,][]{romeo20}. 
In other words, these galaxies appear to be transitioning from a peak starburst phase towards becoming more typical, steadily star-forming galaxies. 

Conversely, at $\Mh \lesssim 10^{11.6} h^{-1}\Msun$, our SHMRs gradually decrease with decreasing halo mass. 
However, the slope at the less-massive end is shallower compared to the prediction from the abundance matching technique presented by \citet{behroozi19}. 
Our observational results suggest that, at $z=1.1-1.4$, the ELGs in less-massive haloes, particularly at $\Mh \lesssim 10^{11} h^{-1}\Msun$, are still undergoing active star formation, although the CGM, which serves as a reservoir of cold gas, is becoming depleted by strong galactic outflows (see Section~\ref{subsubsec:bce}). 
We observe a significant redshift evolution of SHMRs for less-massive ELGs from $z \sim 2.2$ to $z=1.1-1.4$, shifting towards the lower right direction as they assemble their stellar masses and experience more rapid growth of dark halo mass. 
Notably, the SHMRs show no population dependencies for ELGs across all dark halo mass ranges shown in Figure~\ref{fig:shmr}. 

Finally, throughout this subsection, the results for the LHMR, BCE, and SHMR demonstrate that the median values, independently calculated from observed line luminosities, stellar masses, and SFRs, align well with those derived from the posterior means of our HOD-model analysis. 
They also show consistency with trends from previous observational studies. 
This consistency underscores the robustness and reliability of our findings. 
All median values of the physical properties discussed in this section are summarised in Table~\ref{tab:list_median}. 

While our HOD framework is specifically designed for ELGs in this study, its flexibility allows for potential applications to other galaxy populations. 
For instance, the framework could be extended to examine star-forming and passive galaxies or to compare field and cluster galaxies, providing insights into how different galaxy populations and environments influence galaxy luminosity functions and their connection to dark matter haloes. 
Additionally, this framework can be extended to various types of ELGs across a wide range of redshifts, enabling a deeper understanding of the evolution of galaxy--halo connections over cosmic time. 

\begin{table}
\caption{List of the median values of physical properties of our $\oii$-emitting galaxies observed through NB816 and NB921 filters. }
\label{tab:list_median}
\renewcommand{\arraystretch}{1.18}
\begin{tabular}{ccc} \hline \hline
 & NB816 $(z_{\rm eff} = 1.193)$ & NB921 $(z_{\rm eff} = 1.471)$ \\ \hline
$\log_{10}{(M_{\rm min}^{\rm LF}/h^{-1}\Msun)}$ & $11.50^{+0.16}_{-0.16}$ & $11.62^{+0.19}_{-0.19}$ \\
$\log_{10}{(\Mstar/\Msun)}$ & $9.61 \pm 0.42$ & $9.82 \pm 0.45$ \\
$\log_{10}{({\rm SFR/\Msun/yr})}$ & $0.69 \pm 0.38$ & $0.95 \pm 0.47$ \\
$\log_{10}{(L_{\rm ELG}{\rm /erg/s})}$ & $41.65 \pm 0.22$ & $41.94 \pm 0.20$ \\ 
$\log_{10}{({\rm BCE})}$ & $-0.13^{+0.34}_{-0.47}$ & $-0.12^{+0.33}_{-0.38}$ \\
$\log_{10}{({\rm SHMR}/h)}$ & $-2.11^{+0.69}_{-0.70}$ & $-2.01^{+0.83}_{-0.83}$ \\ \hline
\end{tabular}
\end{table}

\section{Summary and Conclusions} \label{sec:conclusion}

In this paper, we have introduced a new HOD model that integrates galaxy luminosity, a key observable reflecting ELG star-formation activity, into the central galaxy occupation function. 
This innovation enables us to predict galaxy LFs from the HOD model and facilitates joint analyses using both ACFs and LFs. 
By representing the differential number density of galaxies as a function of their luminosity, the LFs provide a more accurate representation of galaxy occupation and provide more robust constraints on the HOD parameters compared to models that rely solely on the integrated number density. 

We applied our HOD model to analyse the ACFs and LFs of $\oii$-emitting galaxies at two redshift slices (NB816: $z_{\rm eff} = 1.193$, NB921: $z_{\rm eff} = 1.471$) observed from the HSC SSP PDR2 on the Deep/UD layers \citep{aihara19,hayashi20}. 
We compared the results obtained using our HOD model with those derived from the well-established \citet{geach12} model, widely used for interpreting ELG ACFs. 
Our HOD model successfully reproduced the observed ACFs and LFs, demonstrating its efficacy in deriving host halo characteristics by jointly examining both the spatial distribution and luminous properties of galaxies. 

A comparison between the two models confirms that the physical properties of $\oii$-emitting galaxies, such as ELG number density, effective halo mass, galaxy bias, and satellite fraction, are consistent across both models. 
However, notable differences emerge in the predicted central ELG occupations between the HOD models. 
The Geach model indicates a sharp increase in the expected number of central ELGs at $\Mh \sim 10^{12} h^{-1}\Msun$, whereas our model predicts a more gradual increase in central ELG occupation within haloes ranging from $\Mh \sim 10^{11} h^{-1}\Msun$ to $10^{12} h^{-1}\Msun$ for both the NB816 and NB921 populations.

The best-fitting and posterior mean central occupation functions indicate a minimal contribution from the central ELG occupation with the Gaussian distribution. 
To assess whether this component is necessary, we excluded the Gaussian central ELG occupation from our model and reanalysed the ACFs and LFs, reducing three HOD free parameters. 
We found that the physical properties of ELGs, with or without the Gaussian central occupation in less-massive haloes, remained consistent within $1\sigma$ confidence intervals, suggesting its limited impact on our $\oii$-emitter samples. 

We define the median dark halo mass of central ELGs from the HOD-model analysis as $M_{\rm min}^{\rm LF}$, using it 
to link the observed properties of ELGs to the characteristics of their host dark haloes. 
We predict the halo mass evolution from $M_{\rm min}^{\rm LF}$ for the $\oii$ NB816 and NB921 samples to the present-day Universe using the EPS model, connecting the evolutionary relationship between various galaxy populations and our samples. 
For $\oii$ emitters observed through the NB921 filter at $z\sim1.4$, the $1\sigma$ confidence interval of the evolved dark halo mass at $z=0.0$ fully encompasses the estimated virial mass of the Milky Way, suggesting that these ELGs could be descendants or building blocks of Milky Way-like galaxies in the present-day Universe. 
In contrast, the $\oii$ emitters in the NB816 filter at $z\sim1.2$ are predicted to evolve into galaxies with total masses slightly lower than that of the Milky Way in the local Universe. 

Using the posterior mean results from our HOD model, which excludes the Gaussian central occupation and incorporating the median physical properties derived directly from our ELG samples, we investigated the relationship between central $\oii$ emitters and their host haloes. 
The LHMRs, which describes ELG line luminosities as a double power law relative to halo mass, are consistent with previous observational findings. 
We observe a slight redshift evolution in the LHMRs, indicating that the ELG line luminosities tend to increase with redshift at a fixed dark halo mass. 
However, the LHMRs show no significant differences between ELG populations. 

Our BCEs exhibit a slight excess compared to the model prediction by \citet{behroozi13} at $\Mh \sim 10^{11.6}h^{-1}\Msun$, where the theoretical model predicts the BCEs peak. 
This suggests that the galaxies in our samples are experiencing more active star formation than the average properties at the same cosmic poch. 
Our posterior mean BCEs indicate increasing trends with decreasing halo masses, contrary to the theoretical model's decreasing trend, and these trends align with other observational studies. 
This result supports the occurrence of bursty star formation in ELGs hosted by $\Mh \lesssim 10^{11}h^{-1}\Msun$ haloes. 

Finally, our SHMRs align well with the theoretical predictions for central star-forming galaxies at $z \sim 1.5$ as presented by \citet{behroozi19}. 
This consistency suggests that our $\oii$ emitter samples are approaching the final stages of their rapid stellar mass assembly phase and transitioning from peak starburst activities into more stable states as ordinary star-forming galaxies. 
The slight excess in BCEs compared to the theoretical model further supports this transition, indicating active star formation relative to the average properties at similar cosmic ages. 
This transition highlights the evolving nature of $\oii$ emitters as they integrate into the broader population of star-forming galaxies. 

In conclusion, our proposed HOD model successfully incorporates the luminosity function as a new constraint, thereby enabling a more realistic depiction of how ELGs populate their host dark haloes. 
This advancement not only sheds light on the properties of the underlying dark haloes but also facilitates the generation of high-quality mock catalogues for future extensive surveys. 
Although originally developed for ELGs, the model can be applied to other galaxy populations, provided suitable observational data are available, as the luminosity function is a fundamental observable. 
In forthcoming wide-field surveys such as the Euclid Wide Survey \citep{euclid24}, the High Latitude Wide Area Survey by the Nancy Grace Roman Space Telescope \citep{wang22}, and the Legacy Survey of Space and Time by the Vera C. Rubin Observatory \citep{ivezic19}, verifying that the double power-law LHMR remains valid for more diverse or broad-band-selected galaxy samples will be crucial. 
Hydrodynamical simulations and/or survey mock catalogues will help determine whether the double power-law is sufficient or if additional parameters are needed; if the data indicate a different functional form, refining the HOD model will be pivotal for robustly linking galaxies to their host haloes. 
By applying this framework to other datasets and galaxy populations, future research can extend our understanding of galaxy evolution, refine cosmological models, and further validate the robustness of this approach through dedicated observational campaigns and advanced numerical simulations. 

Incorporating large-volume hydrodynamical simulations such as TNG-Cluster \citep{nelson24} or MillenniumTNG \citep{pakmor23} would offer an excellent test bed for our LHMR, particularly at the bright end where observational data are often sparse. 
These simulations can generate sufficient numbers of rare, high-luminosity ELGs that smaller volumes might fail to capture, thus placing more stringent constraints on the double power-law assumption. 
Furthermore, appropriate dust-attenuation prescriptions in such simulations could help refine how the contamination fraction evolves with halo mass or cosmic time, enabling our model to capture additional population-level dependencies beyond its current scope. 

\section*{Acknowledgements}
The authors thank the anonymous referee for valuable suggestions and comments that significantly improved this manuscript. 
We extend our gratitude to Takahiro Nishimichi, Masahiro Takada, Hironao Miyatake, and Atsushi Taruya for their invaluable advice and insightful discussions throughout this study. 
We also acknowledge Rhythm Shimakawa for sharing the data of observed line flux ratios of Balmer series, and Ken Osato for providing the ELG catalogues from the IllustrisTNG simulations. 
SI acknowledges support from JSPS KAKENHI Grant-in-Aid for Early-Career Scientists (Grant Number 23K13145) and from ISHIZUE 2022 of Kyoto University.
TO acknowledges support of the Taiwan National Science and Technology Council under Grants No. NSTC 112-2112-M-001-034- and NSTC 113-2112-M-001-011-. 
TTT has been supported by the Grants-in-Aid for Scientific Research (24H00247) and the Collaboration Funding of the Institute of Statistical Mathematics ``Machine-Learning-Based Cosmogony: From Structure Formation to Galaxy Evolution''. 

The Hyper Suprime-Cam (HSC) collaboration includes the astronomical communities of Japan and Taiwan, and Princeton University. 
The HSC instrumentation and software were developed by the National Astronomical Observatory of Japan (NAOJ), the Kavli Institute for the Physics and Mathematics of the Universe (Kavli IPMU), the University of Tokyo, the High Energy Accelerator Research Organization (KEK), the Academia Sinica Institute for Astronomy and Astrophysics in Taiwan (ASIAA), and Princeton University. 
Funding was contributed by the FIRST program from the Japanese Cabinet Office, the Ministry of Education, Culture, Sports, Science and Technology (MEXT), the Japan Society for the Promotion of Science (JSPS), Japan Science and Technology Agency (JST), the Toray Science Foundation, NAOJ, Kavli IPMU, KEK, ASIAA, and Princeton University. 

This paper makes use of software developed for Vera C. Rubin Observatory. 
We thank the Rubin Observatory for making their code available as free software at http://pipelines.lsst.io/. 

This paper is based on data collected at the Subaru Telescope and retrieved from the HSC data archive system, which is operated by the Subaru Telescope and Astronomy Data Center (ADC) at NAOJ. 
Data analysis was in part carried out with the cooperation of Center for Computational Astrophysics (CfCA), NAOJ. 
We are honored and grateful for the opportunity of observing the Universe from Maunakea, which has the cultural, historical and natural significance in Hawaii. 

The Pan-STARRS1 Surveys (PS1) and the PS1 public science archive have been made possible through contributions by the Institute for Astronomy, the University of Hawaii, the Pan-STARRS Project Office, the Max Planck Society and its participating institutes, the Max Planck Institute for Astronomy, Heidelberg, and the Max Planck Institute for Extraterrestrial Physics, Garching, The Johns Hopkins University, Durham University, the University of Edinburgh, the Queen's University Belfast, the Harvard-Smithsonian Center for Astrophysics, the Las Cumbres Observatory Global Telescope Network Incorporated, the National Central University of Taiwan, the Space Telescope Science Institute, the National Aeronautics and Space Administration under grant No. NNX08AR22G issued through the Planetary Science Division of the NASA Science Mission Directorate, the National Science Foundation grant No. AST-1238877, the University of Maryland, Eotvos Lorand University (ELTE), the Los Alamos National Laboratory, and the Gordon and Betty Moore Foundation. 

Data analysis was in part carried out on the Multi-wavelength Data Analysis System operated by the Astronomy Data Center (ADC) and the Large-scale data analysis system co-operated by the Astronomy Data Center and Subaru Telescope, National Astronomical Observatory of Japan. 

\section*{Data Availability}
Emission-line galaxy catalogues selected from the Hyper Suprime-Cam Subaru Strategic Programme are available on the {public data release (PDR) webpage}\footnote{https://hsc-release.mtk.nao.ac.jp/doc/}. 



\bibliographystyle{mnras}
\bibliography{hod_with_lf-fit} 




\appendix

\bsp	
\label{lastpage}
\end{document}